\documentclass[twoside,11pt,final]{article}
\usepackage{epsfig}
\usepackage{amssymb}
\usepackage{url}
\usepackage[l]{floatflt}
\usepackage{times,fancyhdr}
\usepackage{geometry}

\usepackage{latexsym}
\usepackage{euscript}
\usepackage{times}
\usepackage{caption}
\usepackage{amsmath}

\newcommand{\ftn}{\footnotesize}

\newcommand{\ssz}{\scriptsize}

\newcommand{\etal}{{\it et al.\/}}

\newcommand{\eec}{\end{center}}
\newcommand{\bec}{\begin{center}}
\newcommand{\eem}{\end{matrix}}
\newcommand{\bem}{\begin{matrix}}
\newcommand{\eeq}{\end{equation}}
\newcommand{\beq}{\begin{equation}}
\newcommand{\ba}{\begin{array}}
\newcommand{\ea}{\end{array}}
\newcommand{\bea}{\begin{eqnarray}}
\newcommand{\eea}{\end{eqnarray}}
\newcommand{\baq}{\begin{eqnarray}}
\newcommand{\eaq}{\end{eqnarray}}
\newcommand{\beqs}{\begin{subequations}}
\newcommand{\eeqs}{\end{subequations}}

\newcommand{\Eref}[1]{Eq.~(\ref{#1})}
\newcommand{\Sref}[1]{Sec.~\ref{#1}}
\newcommand{\Fref}[1]{Fig.~\ref{#1}}
\newcommand{\Tref}[1]{Table~\ref{#1}}
\newcommand{\cref}[1]{Ref.~\cite{#1}}

\newcommand{\sFref}[2]{Fig.~\ref{#1}-{\sf\ftn ({#2})}}
\newcommand{\sTref}[2]{Table~\ref{#1}-{\sf\ftn ({#2})}}

\newcommand\eq[1]{Eq.~(\ref{#1})}
\newcommand\eqs[2]{Eqs.~(\ref{#1}) and (\ref{#2})}
\newcommand\eqss[3]{Eqs.~(\ref{#1}), (\ref{#2}) and (\ref{#3})}

\newcommand{\TeV}{{\mbox{\rm TeV}}}

\newcommand{\GeV}{{\mbox{\rm GeV}}}
\newcommand{\sFig}[2]{Fig.~\ref{#1}-${\sf ({#2})}$}

\def\lf{\left(}
\def\rg{\right)}

\newcommand{\Vhi}{\ensuremath{V_{\rm HI}}}
\newcommand{\chir}{\ensuremath{c_{\rm HI}}}
\newcommand{\Nhi}{\ensuremath{N_{\rm HI*}}}
\newcommand{\Vhio}{\ensuremath{V_{\rm HI0}}}
\newcommand{\ck}{\ensuremath{c_{2K}}}
\newcommand{\ckk}{\ensuremath{c_{4K}}}
\newcommand{\ckx}{\ensuremath{c_{6K}}}
\newcommand{\ckh}{\ensuremath{c_{8K}}}

\newcommand{\kp}{\ensuremath{\kappa}}
\newcommand{\ks}{\ensuremath{k_{4S}}}
\newcommand{\kss}{\ensuremath{k_{6S}}}
\newcommand{\ksss}{\ensuremath{k_{8S}}}
\newcommand{\ksh}{\ensuremath{k_{8S}}}
\newcommand{\kst}{\ensuremath{k_{10S}}}
\newcommand{\ksv}{\ensuremath{k_{12S}}}
\newcommand{\vg}{\ensuremath{v_{_G}}}

\newcommand{\ns}{\ensuremath{n_{\rm s}}}
\newcommand{\as}{\ensuremath{\alpha_{\rm s}}}
\newcommand{\aS}{\ensuremath{{\rm a}_S}}
\newcommand{\Dex}{\ensuremath{\Delta_{\rm m*}}}

\newcommand{\bi}{\ensuremath{\beta_m}}
\newcommand{\al}{\ensuremath{\alpha}}
\newcommand{\bt}{\ensuremath{\beta}}
\newcommand{\hm}{\ensuremath{h_m}}

\newcommand{\phh}{\ensuremath{\Phi}}
\newcommand{\bph}{\ensuremath{\bar \Phi}}
\newcommand{\mP}{\ensuremath{m_{\rm P}}}
\newcommand{\what}{\ensuremath{\widehat}}
\newcommand{\sgm}{\ensuremath{\sigma}}

\def\M{{\bar{M}}}
\def\N{{\bar{N}}}

\def\W{{\what{W}}}
\def\K{{\what{K}}}

\def\Z{\what{Z}}
\def\n{\bar{n}}

\def\Ms{\what{M}_{\rm S}}

\def\Ka{K\"{a}hler potential~}
\def\Kap{K\"{a}hler potential}
\def\ä{\"{a}}
\def\p{|S|}

\renewenvironment{subequations}{%
\refstepcounter{equation}%
\setcounter{parentequation}{\value{equation}}%
  \setcounter{equation}{0}
  \ignorespaces
}{%
  \setcounter{equation}{\value{parentequation}}%
  \ignorespacesafterend
}

\begin{document}

\makeatletter

\def\@maketitle{%
  \newpage
  \null
  \vskip 2em%
  \begin{center}%
  \let \footnote \thanks
    {\LARGE \@title \par}%
    \vskip 1.5em%
    {\large
      \lineskip .5em%
      \begin{tabular}[t]{c}%
        \@author
      \end{tabular}\par}%
    \vskip 1em%
  \end{center}%
  \par
\vskip 2.5em }

\makeatother

\title{\bfseries\scshape Implementing Hilltop F-term Hybrid
\\ Inflation in Supergravity}
\author{{\bfseries\scshape R. Armillis$^{\sf\ftn (1)}$ and C. Pallis$^{\sf\ftn (2)}$}\\ \\
$^{\sf\ftn (1)}${\sl\small Institut de Th\'eorie des Ph\'enomen\`es Physiques,}\\
{\sl\small\'Ecole Polytechnique F\'ed\'erale de
Lausanne,} \\
{\sl\small BSP 730 Cubotron,} {\sl\small  CH-1015 Lausanne, SWITZERLAND}\\
{\tt\ftn roberta.armillis@epfl.ch}\\[0.2cm]
$^{\sf\ftn (2)}${\sl\small Department of Physics,}
{\sl\small University of Cyprus,} \\
{\sl\small P.O. Box 20537, CY-1678 Nicosia, CYPRUS}\\
{\tt\ftn cpallis@ucy.ac.cy}}

\maketitle
\date{\vspace{-5ex}}

\thispagestyle{empty} \setcounter{page}{1}
\thispagestyle{fancy} \fancyhead{} \fancyhead[L]{\sf\small In:
Recent Advances in Cosmology
\\ Editors: A. Travena and B. Soren, pp. {159-192}
}
\fancyhead[R]{\sf\small ISBN 978-1-62417-943-3  \\
\copyright~2013 Nova Science Publishers, Inc.} \fancyfoot{}
\renewcommand{\headrulewidth}{0pt}

\begin{abstract}

{\small F-term hybrid inflation (FHI) of the hilltop type can
generate a scalar spectral index, $n_{\rm s}$, in agreement with
the fitting of the seven-year Wilkinson microwave anisotropy probe
data by the standard power-law cosmological model with cold dark
matter and a cosmological constant, $\Lambda$CDM. We investigate
the realization of this type of FHI by using quasi-canonical
K\"ahler potentials with or without the inclusion of extra
hidden-sector fields. In the first case, acceptable results can be
obtained by constraining the coefficients of the quadratic and/or
quartic supergravity correction to the inflationary potential and
therefore a mild tuning of the relevant term of the K\"ahler
potential is unavoidable. Possible reduction of $n_{\rm s}$
without generating maxima and minima of the potential on the
inflationary path is also possible in a limited region of the
available parameter space. The tuning of the terms of the K\"ahler
potential can be avoided with the adoption of a simple class of
string-inspired K\"ahler potentials for the hidden-sector fields
which ensures a resolution to the $\eta$ problem of FHI and allows
acceptable values for the spectral index, constraining the
coefficient of the quartic supergravity correction to the
inflationary potential. Performing a four-point test of the
analyzed models, we single out the most promising of these. }
\\[0.2cm] {\ssz\sffamily\scshape Keywords: \sf Cosmology, Inflation} \\ {\ssz {\sffamily\scshape PACS
codes:} \sf 98.80.Cq, 11.30.Pb}

\end{abstract}

\newpage
%
\pagestyle{fancy} \fancyhead{} \fancyhead[ER]{\sl R. Armillis \&
C. Pallis} \fancyhead[EL,OR]{\bf \thepage} \fancyhead[OL]{\sl
Implementing Hilltop FHI in SUGRA} \fancyfoot{}
\renewcommand\headrulewidth{0.5pt}



\section{\scshape Prologue}\label{intro}

Inflation \cite{guth} has been incredibly successful in providing
solutions to the problems of the \emph{Standard Big Bang
cosmology} ({\ftn\sf SBB}). It can set the initial conditions,
which give rise to the high degree of flatness and homogeneity
that we observe in the universe today. From particle physics
motivated models, it not only yields a mechanism for accelerated
expansion but also explains, through quantum fluctuations, the
origin of the temperature anisotropies in the \emph{Cosmic
Microwave Background} ({\ftn\sf CMB}) and the seeds for the
observed Large Scale Structure -- for reviews see e.g. Refs.
\cite{review, lectures}.

We focus on a set of well-motivated, popular and quite natural
models of {\it supersymmetric} ({\sf\ftn SUSY}) {\it F-term hybrid
inflation} ({\sf\ftn FHI})~\cite{hybrid}. Namely, we consider the
standard \cite{susyhybrid} FHI and some of its specific versions:
the shifted \cite{jean} and smooth \cite{pana1} FHI. They are
realized~\cite{susyhybrid} at (or close to) the SUSY \emph{Grand
Unified Theory} ({\ftn\sf GUT}) scale $M_{\rm
GUT}\simeq2.86\cdot10^{16}~{\rm GeV}$ and can be easily linked to
several extensions \cite{lectures} of the \emph{Minimal
Supersymmetric Standard Model} ({\ftn\sf MSSM}) which have a rich
structure. Namely, the $\mu$-problem of MSSM is solved via a
direct coupling of the inflaton to Higgs superfields \cite{dvali}
or via a Peccei-Quinn symmetry \cite{rsym}, baryon number
conservation is an automatic consequence \cite{dvali} of an R
symmetry and the baryon asymmetry of the universe is generated via
leptogenesis which takes place \cite{lept} through the
out-of-equilibrium decays of the inflaton's decay products.

Although quite successful, these models have at least two
shortcomings:

\begin{itemize}

\item[\bf (i)] The problem of the enhanced (scalar) spectral
index, $n_{\rm s}$. It is well-known that under the assumption
that the problems of SBB are resolved exclusively by FHI, these
models predict $n_{\rm s}$ just marginally consistent with the
fitting of the seven-year results \cite{wmap} from the
\emph{Wilkinson Microwave Anisotropy Probe Satellite} ({\sf\ftn
WMAP7}) data with the standard power-law cosmological model with
\emph{cold dark matter and a cosmological constant} ({\ftn\sf
$\Lambda$CDM}).

\item[\bf (ii)] The so-called $\eta$ problem. This problem is tied
\cite{review, hybrid, eta} on the expectation that
\emph{supergravity} ({\sf\ftn SUGRA}) corrections generate a mass
squared for the inflaton of the order of the Hubble parameter
during FHI and so, the $\eta$ criterion is generically violated,
ruining thereby FHI. Inclusion of SUGRA corrections with canonical
K\"ahler potential prevents \cite{hybrid, senoguz, sstad} the
generation of such a mass term due to a mutual cancellation.
However, despite its simplicity, the canonical K\"ahler potential
can be regarded \cite{hybrid} as fine tuning to some extent and,
in all cases, increases $n_{\rm s}$ even more.

\end{itemize}

In this topical review we reconsider one set of possible
resolutions (for other proposals, see \cref{battye, mhi, pqhi,
dim}) of the tension between FHI and the data. This is relied on
the utilization of three types of quasi-canonical \cite{CP}
K\"ahler potential with or without the inclusion of extra fields,
$\hm$. The term ``extra fields'' refers to hidden-sector
\cite{martin} fields or fields which do not participate
\cite{sugraP} in the inflationary superpotential but may only
affect the \Kap. The consideration of extra fields assists us in
solving the $\eta$ problem of FHI as well. In particular, we
review the following embeddings of FHI in SUGRA:

\begin{itemize}

\item[\bf (i)]  FHI in \emph{next-to-minimal SUGRA} ({\sf\ftn
nmSUGRA}) -- see \Sref{nmsugra}. A convenient choice of the
next-to-minimal term \cite{gpp,mur,hinova} of the \Ka leads to a
negative mass (quadratic) term for the inflaton and therefore
$\ns$ can be diminished sizeably.

\item[\bf (ii)]  FHI in \emph{next-to-next-to-minimal SUGRA}
({\sf\ftn nnmSUGRA}) -- see \Sref{nnmsugra}. A convenient choice
of the next-to-minimal and the next-to-next-to-minimal term
generates \cite{rlarge,alp} a positive mass (quadratic) term for
the inflaton and a sizeable negative quartic term which yield
acceptable $\ns$ enhancing somehow the running of $\ns$, $\as$.

\item[\bf (iii)] FHI with extra fields, $\hm$, obeying a
string-inspired K\"ahler potential ({\sf\ftn hSUGRA}) -- see
\Sref{hsugra}. In the presence of $\hm$'s, we can establish
\cite{nmhi} a type of FHI which avoids the tuning -- required in
the cases (i) and (ii) above -- of the quadratic SUGRA correction
and is largely dominated by the quartic SUGRA correction. Namely,
the coefficients of the K\"ahler potential are constrained to
natural values (of order unity) so as the mass term of the
inflaton field is identically zero.

\end{itemize}

In all the cases above and in the largest part of the parameter
space the inflationary potential acquires a local maximum and
minimum. Then, FHI of the \emph{hilltop} \cite{lofti, lofti1} type
can occur as the inflaton rolls from this maximum down to smaller
values. However, the value of the inflaton field at the maximum is
to be sufficiently close to the value that this field acquires
when the pivot scale crosses outside the inflationary horizon.
Therefore, $\ns$ can become consistent with data, but only at the
cost of an extra indispensable mild tuning \cite{gpp} of the
initial conditions. Another possible complication is that the
system may get trapped near the minimum of the inflationary
potential, thereby jeopardizing the attainment of FHI. On the
other hand, we can show \cite{mur} that acceptable $n_{\rm s}$'s
can be obtained even maintaining the monotonicity of the
inflationary potential, i.e. without this minimum-maximum problem
in the case of nmSUGRA.

In this presentation we reexamine the above ideas for the
reduction of $n_{\rm s}$ within FHI, updating our results in
\cref{hinova, nmhi} and incorporating recent related developments
in \cref{rlarge}. In particular, the text is organized as follows:
In Sec.~\ref{fhim}, we review the basic FHI models and in
\Sref{msugra} we recall the results holding for FHI in
\emph{minimal SUGRA} ({\sf\ftn mSUGRA}). In the following we
demonstrate how we can obtain hilltop FHI using various types of
K\"ahler potentials -- see Secs~\ref{nmsugra}, \ref{nnmsugra} and
\ref{hsugra}. Our conclusions are summarized in
Sec.~\ref{sec:con}. Throughout the text, charge conjugation is
denoted by a star and brackets are, also, used by applying
disjunctive correspondence.


\section{\scshape FHI within SUGRA}\label{fhim}

We outline the salient features of the basic types of FHI. Namely
we present the relevant superpotentials in Sec~\ref{Winf} and the
SUSY potentials in Sec.~\ref{VFinf}. We then (in
Sec.~\ref{sugra3}) describe  the embedding of these models in
SUGRA and extract the relevant inflationary potential in
Sec.~\ref{Vinf}.

\subsection{\scshape The Relevant Superpotential}\label{Winf}

The F-term hybrid inflation can be realized adopting one of the
superpotentials below:
\begin{equation} \label{Whi} W = \W + W_{\rm FHI}\>\>\mbox{with}\>\>W_{\rm
FHI}=\left\{\bem
\what\kappa S\left(\bar \Phi\Phi-M^2\right)\hfill   & \mbox{for
standard FHI}, \hfill \cr
\what\kappa S\left(\bar \Phi\Phi-M^2\right)-S{(\bar
\Phi\Phi)^2\over M_{\rm S}^2}\hfill  &\mbox{for shifted FHI},
\hfill \cr
S\left({(\bar \Phi\Phi)^2\over M_{\rm S}^2}-\what\mu_{\rm
S}^2\right)\hfill  &\mbox{for smooth FHI}, \hfill \cr\eem
\right. \end{equation}
where we allow for the presence of a part, $\W$, which depends
exclusively on the hidden sector superfields, $\hm$. Here (and
hereafter) we use the hat to denote such quantities. To keep our
analysis as general as possible, we do not adopt any particular
form for $\W$ -- for some proposals see \cref{covi,davis}. Note
that our construction remains intact even if we set~$\W=0$ as it
was supposed in \cref{sugraP}. This is due to the fact that $\W$
is expected to be much smaller than the inflationary energy
density -- see \Sref{sugra3}. For $\W\neq0$, though, we need to
assume that $h_m$'s are stabilized before the onset of FHI by some
mechanism not consistently taken into account here. As a
consequence, we neglect the dependence of~$\W$, $\what\kappa$ and
$\what\mu_{\rm S}$ on $h_m$ and so, these quantities are treated
as constants. We further assume that the D-terms due to $h_m$'s
vanish -- contrary to the strategy adopted in \cref{sugraP}.

The remaining symbols in the \emph{right hand side} (r.h.s) of
\Eref{Whi} are identified as follows:

\begin{itemize}

\item $S$ is a left handed superfield, singlet under a GUT gauge
group $G$;

\item $\bar{\Phi}$, $\Phi$ is a  pair of left handed superfields
belonging to non-trivial conjugate representations of $G$, and
reducing its rank by their {\it vacuum expectation values}
(v.e.vs);

\item $\Ms\sim 5\times10^{17}~{\rm GeV}$ is an effective cutoff
scale comparable with the string scale;

\item $\what\kappa$ and $\what M,~\what\mu_{\rm S}~(\sim M_{\rm
GUT})$ are parameters which can be made positive by field
redefinitions.

\end{itemize}

The superpotential in Eq.~(\ref{Whi}) for standard FHI is the most
general renormalizable superpotential consistent with a continuous
R-symmetry \cite{susyhybrid} under which
\begin{equation}
  \label{Rsym}
S\  \rightarrow\ e^{ir}\,S,~\bar\Phi\Phi\ \rightarrow\
\bar\Phi\Phi,~W \rightarrow\ e^{ir}\, W.
\end{equation}
After including in this superpotential the leading
non-renormalizable term, one obtains the superpotential of shifted
\cite{jean} FHI in Eq.~(\ref{Whi}). Finally, the superpotential of
smooth \cite{pana1} FHI can be obtained if we impose an extra
$Z_2$ symmetry under which the combination $\bar{\Phi}\Phi$ has
unit charge.

\subsection{\scshape The SUSY Potential}\label{VFinf}

The SUSY potential, $V_{\rm SUSY}$, extracted (see e.g.
ref.~\cite{review}) from $W_{\rm FHI}$ in Eq.~(\ref{Whi}) includes
F and D-term contributions. Namely,
\beq V_{\rm SUSY}=V_{\rm F}+V_{\rm D},\label{Vsusy}\eeq
where

\paragraph{\bf (i)} The F-term contribution can be written as:
\beq \label{VF} V_{\rm F}=\left\{\bem
\kappa^2M^4\left(({\sf\ftn \Phi}^2-1)^2+2{\sf\ftn S}^2{\sf\ftn
\Phi}^2\right)\hfill & \mbox{for standard FHI}, \hfill \cr
\kappa^2M^4\left(({\sf\ftn \Phi}^2-1-\xi{\sf\ftn
\Phi}^4)^2+2{\sf\ftn S}^2{\sf\ftn \Phi}^2(1-2\xi{\sf\ftn
\Phi}^2)^2\right)\hfill &\mbox{for shifted FHI}, \hfill \cr
\mu^4_{\rm S}\left((1-{\sf\ftn \Phi}^{4})^2+8{\sf\ftn S}^2{\sf\ftn
\Phi}^{6}\right) \hfill &\mbox{for smooth FHI}, \hfill \cr\eem
\right.\eeq
where the scalar components of the superfields are denoted by the
same symbols as the corresponding superfields and $\xi=M^2/\kappa
M^2_{\rm S}$ with $4<1/\xi<7.2$ \cite{jean}. In order to recover
the properly normalized energy density during FHI, we absorb in
the constants of \eq{VF} some normalization pre-factors emerging
from the SUGRA potential $V_{\rm SUGRA}$ -- see below -- so that
their definition is
\beq \label{hatted}\kappa=e^{\K/2\mP^2}\Z^{-1/2}\what\kappa,~~
\mu_{\rm S}=e^{\K/4\mP^2}\Z^{-1/4}\what\mu_{\rm S}~~\mbox{and}~~
M_{\rm S}=e^{-\K/4\mP^2}\Z^{1/4}\Ms\eeq
where $\K$ and $\Z$ are the $h_m$-dependent parts of the \Kap,
$K$, considered in \Sref{sugra3}. The last relation is introduced
so as $\kappa M^2_{\rm S}=\what\kappa \what M^2_{\rm S}$ and
$\mu_{\rm S} M_{\rm S}=\what\mu_{\rm S}\Ms$. Also, we use
\cite{jean, pana1} the following dimensionless quantities
\beq\left\{\bem
{\sf\ftn \Phi}=|\Phi|/M~~\mbox{and}~~{\sf\ftn
S}=\Z^{1/2}|S|/M\hfill & \mbox{for standard or shifted FHI,}
\hfill\cr
{\sf\ftn \Phi}=|\Phi|/\sqrt{\mu_{\rm S} M_{\rm
S}}~~\mbox{and}~~{\sf\ftn S}=\Z^{1/2}|S|/\sqrt{\mu_{\rm S} M_{\rm
S}}\hfill &\mbox{for smooth FHI.} \hfill \cr\eem
\right.\eeq

\begin{figure}[t]\vspace*{-.19in}
\begin{minipage}{8in}
\epsfig{file=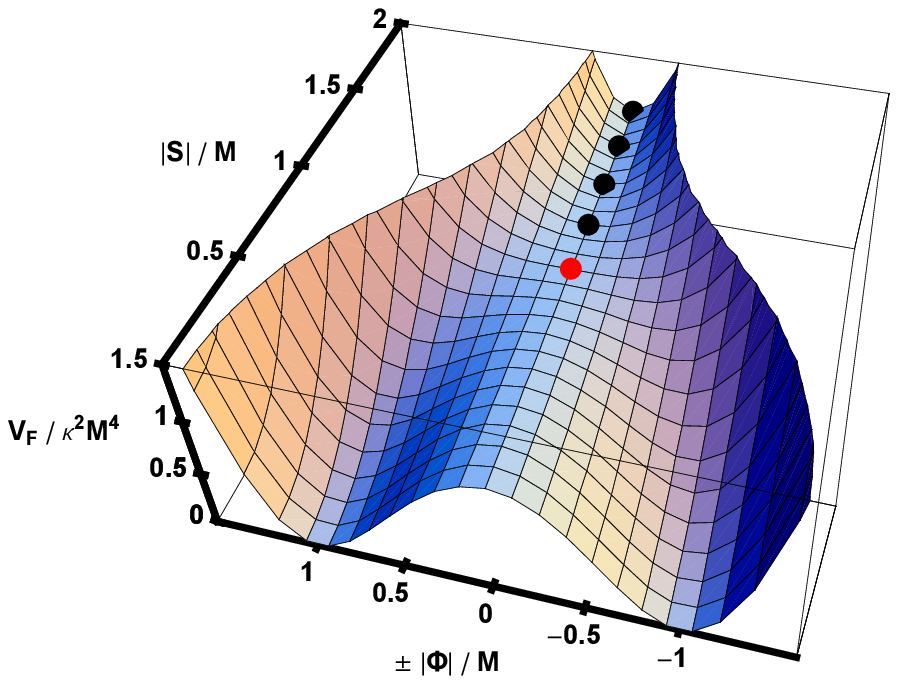,width=7.2cm,,angle=-0}
\hspace*{-0.6cm}
\epsfig{file=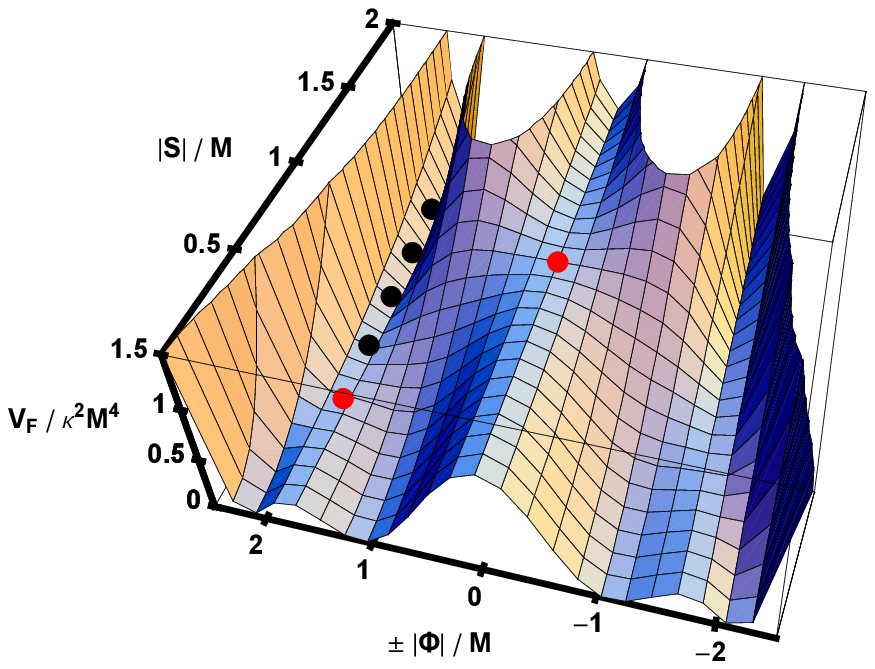,width=7.2cm,,angle=-0} \hfill
\end{minipage}
\hfill \caption[]{\sl\ftn The three dimensional plot of the
(dimensionless) F-term potential $V_{\rm F}/\kappa^2 M^4$ versus
${\sf\ftn S}=|S|/M~~\mbox{and}~~\pm{\sf\ftn \Phi}=\pm|\Phi|/M$ for
standard [shifted with $\xi=1/6$] FHI (left [right]). The
inflationary trajectories are also depicted by black points
whereas the critical points (of the shifted and standard
trajectories) are depicted by red/light points.} \label{Vstad}
\end{figure}

In \sFref{Vstad}{a}, \sFref{Vstad}{b} and  \ref{Vsm} we present
the three dimensional plot of $V_{\rm F}$ versus $\pm{\sf\ftn
\Phi}$ and ${\sf\ftn S}$ for standard, shifted and smooth FHI,
respectively.
\paragraph{\bf (ii)} The D-term contribution $V_{\rm D}$ vanishes for
$\vert\bar{\Phi} \vert=\vert\Phi\vert$ since $V_{\rm D}$ has the
form:
\beq V_{\rm D}= {1\over2}g^2
\sum_aD_aD_a~~\mbox{with}~~D_a=\phi_M\lf T_a\rg^M_\N
K^\N\simeq|\Phi|^2-|\bar\Phi|^2, \eeq
where $g$ is the (unified) gauge coupling constant, $T_a$ are the
generators of $G$ and the notation used is explained below
\Eref{Vsugra} -- recall that $\Phi$ and $\bar\Phi$ belong to the
conjugate representation of $G$.

\paragraph{} From the form of $V_{\rm SUSY}$ in \Eref{VF}, we can understand that $W$ in Eq.~(\ref{Whi})
plays a twofold crucial role:

\paragraph{\bf (i)} It leads to the spontaneous breaking of $G$. Indeed, the
vanishing of $V_{\rm F}$ gives the v.e.vs of the fields in the
SUSY vacuum. Namely,
\begin{equation} \label{vevs} \langle S\rangle=0~~\mbox{and}~~\vert\langle\bar{\Phi}
\rangle\vert=\vert\langle\Phi\rangle\vert=v_{_G}=\left\{\bem
M\hfill   & \mbox{for standard FHI}, \hfill \cr
\frac{M\sqrt{1-\sqrt{1-4\xi}}}{\sqrt{2\xi}}\hfill  &\mbox{for
shifted FHI}, \hfill \cr
\sqrt{\mu_{\rm S}M_{\rm S}}\hfill  &\mbox{for smooth FHI} \hfill
\cr\eem
\right. \end{equation}
(in the case where $\bar{\Phi}$, $\Phi$ are not {\it Standard
Model} (SM) singlets, $\langle\bar{\Phi} \rangle$, $\langle{\Phi}
\rangle$ stand for the v.e.vs of their SM singlet directions). The
non-zero value of the v.e.v $v_{_G}$ signalizes the spontaneous
breaking of $G$.

\paragraph{\bf (ii)} It gives rise to FHI. This is due to the fact that, for
large enough values of $|S|$, there exist valleys of local minima
of the classical potential with constant (or almost constant in
the case of smooth FHI) values of $V_{\rm F}$. In particular, we
can observe that $V_{\rm F}$ takes the following constant value
\beq V_{\rm HI0}=\left\{\bem
\kappa^2 M^4\hfill \cr
\kappa^2 M_\xi^4\hfill \cr
\mu_{\rm S}^4\hfill \cr\eem
\right.\>\>\mbox{along the direction(s):}\>\>{\sf\ftn
\Phi}=\left\{\bem
0~~\hfill & \mbox{for standard FHI}, \hfill\cr
0\>\>\mbox{or}\>\>1/\sqrt{2\xi} \hfill &\mbox{for shifted FHI},
\hfill \cr
0\>\>\mbox{or}\>\>1/2\sqrt{3}{\sf\ftn S} \hfill &\mbox{for smooth
FHI}, \hfill \cr\eem
\right.\eeq
with $M_\xi=M\sqrt{1/4\xi-1}$. From Figs.~\ref{Vstad} and
\ref{Vsm} we deduce that the flat direction ${\sf\ftn \Phi}=0$
corresponds to a minimum of $V_{\rm F}$, for $|S|\gg M$, in the
cases of standard and shifted FHI and to a maximum of $V_{\rm F}$
in the case of smooth FHI. The inflationary trajectories are
depicted by bold points, whereas the critical points by red/light
points. Note that critical points exist only in the case of
standard -- for ${\sf\ftn S}=1$ -- and shifted -- for ${\sf\ftn
S}=(1/4\xi-1)/2$ -- FHI but not for smooth FHI. In the case of
\sFref{Vstad}{b}, the implementation of shifted FHI is ensured by
restricting $1/\xi$ in the range $(4-7.2)$ \cite{jean}. Under this
assumption, the shifted track lies lower that the trivial one and
so, it is energetically more favorable to drive FHI.

Since FHI can be attained along a minimum of $V_{\rm F}$, we infer
that, during standard FHI, the GUT gauge group $G$ is necessarily
restored. As a consequence, topological defects such as strings
\cite{mairi, gpp}, monopoles, or domain walls may be produced
\cite{pana1} via the Kibble mechanism \cite{kibble} during the
spontaneous breaking of $G$ at the end of FHI. This can be avoided
in the other two cases, since the form of $V_{\rm F}$ allows for
non-trivial inflationary valleys along which $G$ is spontaneously
broken due to non-zero values that $\bar \Phi$ and $\Phi$ acquire
during FHI. Therefore, no topological defects are produced in
these cases.

%
 \begin{center}
\begin{minipage}{0.6\textwidth}
\centerline{\epsfig{file=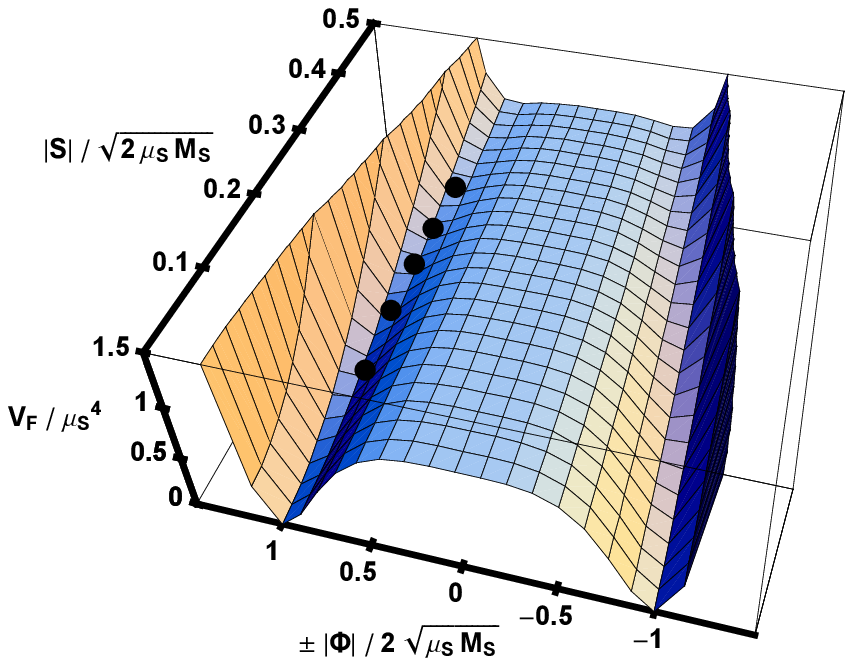,angle=-0,width=7.2cm}}
\hfill
\end{minipage}\hfill
\begin{minipage}{0.4\textwidth}
\captionof{figure}{\sl\ftn The three dimensional plot of the
(dimensionless) F-term potential $V_{\rm F}/\mu_{\rm S}^4$ for
smooth FHI versus ${\sf\ftn S}=|S|/\sqrt{\mu_{\rm S} M_{\rm
S}}~~\mbox{and}~~\pm{\sf\ftn \Phi}=\pm|\Phi|/\sqrt{\mu_{\rm S}
M_{\rm S}}$. The inflationary trajectory is also depicted by black
points.}\label{Vsm}
\end{minipage}
\end{center}

\subsection{\scshape SUGRA Corrections}\label{sugra3}

The consequences that SUGRA has on the models of FHI can be
investigated by restricting ourselves to the inflationary
trajectory $\phh=\bph\simeq0$ (possible corrections due to the
non-vanishing $\phh$ and $\bph$ in the cases of shifted and smooth
FHI are expected to be negligible). Therefore, $W$ in \Eref{Whi}
takes the form
\beq\label{W} W = \W + I\ ,\>\>\mbox{where}\>\>I=-\what V_{\rm
HI0}^{1/2}S\>\>\mbox{with}\>\>\what V_{\rm
HI0}=e^{-\K/\mP^2}\Z\,V_{\rm HI0}. \eeq
The SUGRA scalar potential (without the D-terms) is given (see,
e.g., \cref{review}) by
\begin{equation}
V_{\rm SUGRA}=e^{K\over\mP^2}\left(K^{M\N}F_M\, F^*_\N -3{\vert
W\vert^2\over\mP^2}\right)\>\>\mbox{where}\>\>F_N=W_N +K_N
{W\over\mP^2} \label{Vsugra}
\end{equation}
are the SUGRA-generalized F-terms, the subscript $M~[\M]$ denotes
derivation \emph{with respect to} ({\sf\ftn w.r.t}) the complex
scalar field $\phi_M~[\phi_M^{*}]$ which corresponds to the chiral
superfield $\phi_M$ with $\phi_M=h_m,S,\phh,\bph$ and the matrix
$K^{M\N}$ is the inverse of the K\"ahler metric $K_{M\N}$. In this
paper we consider a quite generic form of K\"{a}hler potentials,
which do not deviate much from the canonical one and respect the R
symmetry of \Eref{Rsym}. Namely we take
\bea\nonumber
K&=&\K+\Z\p^2+{1\over4}\ks\Z^2{\p^4\over\mP^2}+{1\over6}\kss\Z^3{\p^6\over\mP^4}
+{1\over8}\ksss\Z^4{\p^8\over\mP^6} \\ \label{K}
&&+{1\over10}\kst\Z^5{\p^{10}\over\mP^8}
+{1\over12}\ksv\Z^6{\p^{12}\over\mP^{10}}+|\phh|^2+|\bph|^2+\cdots,
\eea
where $\ks,\kss,\ksss,\kst$ and $\ksv$ are positive or negative
constants of order unity and the ellipsis represents higher order
terms involving the waterfall fields ($\Phi$ and $\bar \Phi$) and
$S$. We can neglect these terms since they are irrelevant along
the inflationary path.

Substituting Eq.~(\ref{K}) into \Eref{Vsugra} and expanding
$V_{\rm SUGRA}$ in powers of $\p$, we end up with an expansion of
the form:
\beq \label{Vsugra1} V_{\rm
SUGRA}\simeq\Vhio\lf1-2c_{1K}e^{-{\K\over2\mP^2}}{|\W|\Z^{1/2}\over\sqrt{\Vhio}}\p\cos\theta
+\sum_{\nu=1}^{5}(-1)^{\nu}c_{2\nu K}\Z^\nu\lf
|S|\over\mP\rg^{2\nu}\rg\eeq
where the phase $\theta$ reads $\theta=\arg\W+\arg S+\arg\what
V_{\rm HI0}^{1/2}$. In the r.h.s of the expression above, we
neglect terms proportional to $|\W|^2$ which are certainly
subdominant compared with those which are proportional to $\Vhi$.
From the terms proportional to $|\W|\Vhio^{1/2}$ we present the
second term of the r.h.s of \Eref{Vsugra1} which expresses the
most important contribution \cite{sstad, covi} to the inflationary
potential from the soft SUSY breaking terms. For natural values of
$\W$ and $e^{\K/2\mP^2}$ this term starts \cite{sstad} playing an
important role in the case of standard FHI in mSUGRA  -- see
\Sref{msugra2} -- for $\kappa\lesssim5\cdot10^{-4}$ whereas it has
\cite{sstad} no significant effect in the cases of shifted and
smooth FHI.

Taking in \Eref{K} $\K=0$ and $\Z=1$ (as in Secs.~\ref{msugra} --
\ref{nnmsugra}) the coefficients $c_{\nu K}=c^{(0)}_{\nu K}$ are
found to be
\beqs\bea \label{c1k} c^{(0)}_{1K}&=&-2,\\ \label{c2k}
c^{(0)}_{2K}&=&\ks,\\  \label{c4k}
c^{(0)}_{4K}&=&{1\over2} - {7 \ks\over4} + \ks^2 - {3\kss\over2},\\
c^{(0)}_{6K}&=&-{2\over3} + {3 \ks\over2} - {7 \ks^2\over4} +
\ks^3 + {10\kss\over3} - 3 \ks\kss + 2\ksss,\label{c6k}\\
c^{(0)}_{8K}&=& {3\over8} - {5 \kst\over2} - {13 \ks\over24} + {41
\ks^2\over32} - {7 \ks^3\over4} + \ks^4 - {13\kss\over4}\nonumber
\\&& + {143 \ks\kss\over24} - {9 \ks^2\kss\over2} +
{9\kss^2\over4} - {39\ksss\over8} + 4 \ks \ksss,
\label{c8k}\\c^{(0)}_{10K}&=&-{2\over15} + {32 \kst\over5} + 3
\ksv + {\ks\over24} - 5 \kst \ks - {13 \ks^2\over24} + {41
\ks^3\over32} \nonumber \\ && - {7 \ks^4\over4} + \ks^5 +
{5\kss\over3} - {29 \ks\kss\over6} + {103 \ks^2\kss\over12} - 6
\ks^3\kss - 5\kss^2 \nonumber \\ && + {27 \ks \kss^2\over4} +
5\ksss - {67 \ks\ksss\over8} + 6 \ks^2\ksss- 6\kss \ksss.
\label{c10k}\eea\eeqs
We observe that terms of order $\nu$ in the expansion of \Eref{K}
give rise to contributions of order equal or greater than
$(\nu-2)$ in the expansion of \Eref{Vsugra1}.

\subsection{\scshape The Inflationary Potential} \label{Vinf}

The general form of the potential which can drive the various
versions of FHI reads
\beq\label{Vol} V_{\rm HI}\simeq V_{\rm HI0}\left(1+c_{\rm
HI}-\aS{\sgm\over\sqrt{2\Vhio}}+\sum_{\nu=1}^5(-1)^{\nu}c_{2\nu
K}\lf{\sgm\over\sqrt{2}\mP}\rg^{2\nu}\right),\eeq
where $\sigma=\sqrt{2}\Z^{1/2}\p$ is the canonically (up to the
order $\p^2$) normalized inflaton field and we take $\theta=\pi$
which minimizes $V_{\rm SUGRA}$ for given $\sgm$. To facilitate
our numerical analysis, we introduce the real tadpole parameter
${\rm a}_S$ defined in terms of the $V_{\rm SUGRA}$ parameter, by
the relation
\beq \aS=2c_{1K}e^{-{\K/2\mP^2}}|\W|. \eeq
In \Eref{Vol}, besides the contributions originating from $V_{\rm
SUGRA}$ in \eq{Vsugra1}, we include the term $c_{\rm HI}V_{\rm
HI0}$ which represents a correction to $V_{\rm HI}$ resulting from
the SUSY breaking on the inflationary valley, in the cases of
standard \cite{susyhybrid} and shifted \cite{jean} FHI, or from
the structure of the classical potential in the case of smooth
\cite{pana1} FHI. Indeed, $V_{\rm HI0}>0$ breaks SUSY and gives
rise to logarithmic radiative corrections to the potential
originating from a mass splitting in the $\Phi-\bar{\Phi}$
supermultiplets. On the other hand, in the case of smooth
\cite{pana1} FHI, the inflationary valleys are not classically
flat and, thus, the radiative corrections are expected to be
subdominant. The term $c_{\rm HI}$ can be written as follows:
\begin{equation} \label{Vcor} c_{\rm HI}=\left\{\bem
{\kappa^2{\sf\ftn N}}\left[2 \ln\left(\kappa^2x M^2 /
Q^2\right)+f_{\rm rc}(x)\right]/32\pi^2\hfill
  & \mbox{for standard FHI}, \hfill \cr
{\kappa^2}\left[2 \ln\left(\kappa^2x_\xi M_\xi^2
/Q^2\right)+f_{\rm rc}(x_\xi)\right]/16\pi^2\hfill &\mbox{for
shifted FHI,} \hfill \cr
-2\mu_{\rm s}^2M_{\rm S}^2/27\sigma^4\hfill  &\mbox{for smooth
FHI}, \hfill \cr\eem
\right.\end{equation}
with $x=\sigma^2/2M^2,~x_\xi=\sigma^2/M^2_\xi$ and \beq
\label{frc1} f_{\rm
rc}(x)=(x+1)^{2}\ln(1+1/x)+(x-1)^{2}\ln(1-1/x)\Rightarrow f_{\rm
rc}(x)\simeq3\>\>\mbox{for}\>\>x\gg1.\eeq
Also ${\sf\ftn N}$ is the dimensionality of the representations to
which $\bar{\Phi}$ and $\Phi$ belong and $Q$ is a renormalization
scale. For the values of $\kappa$ encountered in our work
renormalization group effects \cite{espinoza} remain negligible.

In our applications in Secs.~\ref{msugra2}, \ref{nmsugra3},
\ref{nnmsugra3} and \ref{hsugra3} we take ${\sf\ftn N} = 2$. This
choice corresponds to the left-right symmetric GUT gauge group
$SU(3)_c\times SU(2)_L \times SU(2)_R \times U(1)_{B-L}$ with
$\bar\Phi$ and $\Phi$ belonging to $SU(2)_R$ doublets with $B - L
= -1$ and 1 respectively. No cosmic strings are produced during
the GUT phase transition and, consequently, no extra restrictions
on the parameters (as e.g. in Refs. \cite{mairi}) have to be
imposed. As regards the case of shifted \cite{jean} FHI we
identify $G$ with the Pati-Salam gauge group $SU(4)_c \times
SU(2)_L \times SU(2)_R$. Needless to say that the case of smooth
FHI is independent on the adopted GUT since the inclination of the
inflationary path is generated at the classical level and the
addition of any radiative correction is expected to be
subdominant. Negligible is also the third term in the r.h.s of
\Eref{Vol} for $\aS\sim1~\TeV$, besides the case of standard FHI
in mSUGRA -- see \Sref{msugra} -- where it may be important for
$\kappa\leq5\cdot10^{-4}$. For simplicity, we neglect it, in the
analysis of the remaining cases -- see Secs.~\ref{nmsugra},
\ref{nnmsugra} and \ref{hsugra}.

\section{\scshape Constraining FHI}\label{cons}

The parameters of FHI models can be restricted imposing a number
of observational constraints described in Secs.~\ref{obs1} and
\ref{obs2}. Additional theoretical considerations presented in
\Sref{obs3} can impose further limitations.

\subsection{\scshape Inflationary Observables}\label{obs1}

Applying standard  formulae -- see e.g. Refs.~\cite{review,
lectures} -- we can estimate the inflationary observables of FHI.
Namely, we can find:

\paragraph{\bf (i)} The number of e-foldings $N_{\rm HI*}$ that the scale
$k_*=0.002/{\rm Mpc}$ suffers during FHI,
\begin{equation} \label{Nefold}
 N_{\rm HI*}=\:\frac{1}{m^2_{\rm P}}\;
\int_{\sigma_{\rm f}}^{\sigma_{*}}\, d\sigma\: \frac{V_{\rm
HI}}{V'_{\rm HI}},
\end{equation}
where the prime denotes derivation w.r.t $\sigma$, $\sigma_{*}$ is
the value of $\sigma$ when the scale $k_*$ crosses outside the
horizon of FHI, and $\sigma_{\rm f}$ is the value of $\sigma$ at
the end of FHI, which can be found, in the slow roll
approximation, from the condition
\beq \label{slow} {\sf\ftn max}\{\epsilon(\sigma_{\rm
f}),|\eta(\sigma_{\rm f})|\}=1,~~\mbox{where}~~
\epsilon\simeq{m^2_{\rm P}\over2}\left(\frac{V'_{\rm HI}}{V_{\rm
HI}}\right)^2~~\mbox{and}~~\eta\simeq m^2_{\rm P}~\frac{V''_{\rm
HI}}{V_{\rm HI}}\cdot \eeq
In the cases of standard \cite{susyhybrid} and shifted \cite{jean}
FHI and in the parameter space where the terms in
Eq.~(\ref{Vsugra}) do not play an important role, the end of
inflation coincides with the onset of the GUT phase transition,
i.e. the slow roll conditions are violated close to the critical
point $\sigma_{\rm c}=\sqrt{2}M$ [$\sigma_{\rm c}=M_\xi$] for
standard [shifted] FHI, where the waterfall regime commences. On
the contrary,  the end of smooth \cite{pana1} FHI is not abrupt
since the inflationary path is stable w.r.t $\Phi-\bar \Phi$ for
all $\sigma$'s and $\sigma_{\rm f}$ is found from
Eq.~(\ref{slow}). An accurate enough estimation of $\sigma_{\rm
f}$'s -- suitable for our analytical expressions presented below
-- is
\begin{equation} \label{sigmaf} \sigma_{\rm f}\simeq\left\{\bem
\sqrt{2}M\hfill   & \mbox{for standard FHI}, \hfill \cr
M_\xi\hfill &\mbox{for shifted FHI}, \hfill \cr
\sqrt{2\sqrt[3]{5}/3} \sqrt[3]{\mu_{\rm S} M_{\rm S} \mP}\hfill
&\mbox{for smooth FHI}. \hfill \cr\eem
\right.\end{equation}

\paragraph{\bf (ii)} The power spectrum $\Delta^2_{\cal R}$ of the curvature
perturbations generated by $\sigma$ at the pivot scale $k_{*}$
\begin{equation}  \label{Pr}
\Delta_{\cal R}=\: \frac{1}{2\sqrt{3}\, \pi m^3_{\rm P}}\;
\left.\frac{V_{\rm HI}^{3/2}}{|V'_{\rm
HI}|}\right\vert_{\sigma=\sigma_*}\cdot
\end{equation}

\paragraph{\bf (iii)}  The spectral index
\beqs\beq \label{nS}  n_{\rm s}=1-6\epsilon_*\ +\ 2\eta_*, \eeq
its running
\beq \label{aS} \alpha_{\rm s}={2\over3}\left(4\eta_*^2-(n_{\rm
s}-1)^2\right)-2\xi_*,\eeq
with $\xi\simeq m_{\rm P}^4~V'_{\rm HI} V'''_{\rm HI}/V^2_{\rm
HI}$ and the scalar-to-tensor ratio \beq r=16\epsilon_* \eeq\eeqs
where all the variables with the subscript $*$ are evaluated at
$\sigma=\sigma_{*}$.

\subsection{\scshape Observational Constraints}\label{obs2}

Under the assumption that the contribution in Eq.~(\ref{Pr}) is
solely responsible for the observed curvature perturbation  --
i.e. there are no contributions to $\Delta_{\cal R}$ from
curvatons \cite{review} or topological defects \cite{mairi} -- and
(ii) there is a conventional cosmological evolution after FHI --
see point (i) below --, the parameters of the FHI models can be
restricted imposing the following requirements:

\paragraph{\bf (i)} The number of e-foldings $N_{\rm HI*}$
computed by means of \Eref{Nefold}  has to be set equal to the
number of e-foldings $N_*$ elapsed between the horizon crossing of
the observationally relevant mode $k_*$ and the end of FHI. $N_*$
can be found as follows \cite{review}:
\bea \nonumber \frac{k_*}{H_0R_0}=\frac{H_*R_*}{H_0R_0}
&=&\frac{H_*}{H_0}\frac{R_*}{R_{\rm Hf}} \frac{R_{\rm Hf}}{R_{\rm
rh}}\frac{R_{\rm rh}}{R_{\rm eq}}\frac{R_{\rm eq}}{R_0}\\\nonumber
&=& \sqrt{V_{\rm HI0}\over{\rho_{\rm c0} }}e^{-N_*}\left({V_{\rm
HI0}\over\rho_{\rm rh}}\right)^{-1/3}\left({\rho_{\rm
rh}\over\rho_{\rm eq}}\right)^{-1/4}\left({\rho_{\rm
eq}\over\rho_{\rm m0}}\right)^{-1/3}\\ & \Rightarrow&
N_*\simeq\ln{H_0R_0\over k_*} +24.72+{2\over 3}\ln{V^{1/4}_{\rm
HI0}\over{1~{\rm GeV}}}+ {1\over3}\ln {T_{\rm rh}\over{1~{\rm
GeV}}},\label{hor3} \eea
where $T_{\rm rh}$ is the reheating temperature after the
completion of the FHI. Moreover, $R$ is the scale factor, $H=\dot
R/R$ is the Hubble rate, $\rho$ is the energy density and the
subscripts $0$, $k$, Hf, rh, eq and m denote respectively values
at the present (except for the symbol $V_{\rm HI0}$), at the
horizon crossing ($k=R_kH_k$) of the mode $k$, at the end of FHI,
at the end of the reheating period, at the radiation-matter
equidensity point and in the matter dominated era. In our
calculation we take into account that $R\propto \rho^{-1/3}$ for
decaying-particle domination or matter dominated era and $R\propto
\rho^{-1/4}$ for radiation dominated era. We use the following
numerical values:
\beqs\bea &&\rho_{\rm c0}=8.099\times10^{-47}h_0^2~{\rm GeV^4
}~~\mbox{with}~~h_0=0.71,\\&& \rho_{\rm
rh}={\pi^2\over30}g_{\rho*}T_{\rm
rh}^4~~\mbox{with}~~g_{\rho*}=228.75,
\\ && \rho_{\rm eq}=2\Omega_{\rm m0}(1-z_{\rm eq})^3\rho_{\rm
c0}~~\mbox{with}~~\Omega_{\rm m0}=0.26~~\mbox{and}~~z_{\rm
eq}=3135. \eea\eeqs
Setting $H_0=2.37\cdot10^{-4}/{\rm Mpc}$ and $k/R_0=0.002/{\rm
Mpc}$ in Eq.~(\ref{hor3}) we arrive at
\begin{equation}  \label{Nhi}
N_{\rm HI*}\simeq22.6+{2\over 3}\ln{V^{1/4}_{\rm HI0}\over{1~{\rm
GeV}}}+ {1\over3}\ln {T_{\rm rh}\over{1~{\rm GeV}}}\cdot
\end{equation}
Throughout our investigation we take $T_{\rm rh}\simeq10^{9}~\GeV$
as in the majority of these models \cite{lectures,lept, sstad}
saturating conservatively the gravitino constraint
\cite{gravitino}. This choice for $T_{\rm rh}$ does not affect
crucially our results, since $T_{\rm rh}$ appears in
Eq.~(\ref{Nhi}) under its logarithm raised to the one third power
and therefore, its variation over two or three orders of magnitude
has a minor influence on the final value of $\Nhi$.

\paragraph{\bf (ii)} The power spectrum of the curvature perturbations
given by Eq.~(\ref{Pr}) is to be confronted with the WMAP7
data~\cite{wmap}:
\begin{equation}  \label{Prob} \Delta_{\cal R}\simeq\: 4.93\cdot
10^{-5}~~\mbox{at}~~k_*=0.002/{\rm Mpc}.
\end{equation}

\paragraph{\bf (iii)}  According to the
fitting of the WMAP7 results by the cosmological model
$\Lambda$CDM, $n_{\rm s}$ at the pivot scale $k_*=0.002/{\rm Mpc}$
has to fall within the following range of values \cite{wmap}:
\begin{equation}\label{nswmap}
n_{\rm s}=0.968\pm0.024~\Rightarrow~0.944\lesssim n_{\rm s}
\lesssim 0.992~~\mbox{at 95$\%$ c.l.}
\end{equation}

\paragraph{\bf (iv)} Limiting ourselves to $a_{\rm s}$'s consistent with the assumptions of
the power-law $\Lambda$CDM cosmological model, we have to ensure
that $|a_{\rm s}|$ remains negligible. Since, within the
cosmological models with running spectral index, $|a_{\rm s}|$'s
of order 0.01 are encountered \cite{wmap}, we impose the following
upper bound: \beq |a_{\rm s}|\ll0.01.\label{aswmap}\eeq

\subsection{\scshape Theoretical Considerations}\label{obs3}

From a more theoretical point of view, the models of (hilltop) FHI
can be better refined using the following criteria:

\paragraph{\bf (i)}  Gauge coupling unification. When $G$ contains non-abelian
factors (beyond the SM one), the mass, $g\vg$, of the lightest
gauge boson at the SUSY vacuum, \Eref{vevs} is to take the value
dictated by the unification of the gauge coupling constants within
MSSM, i.e.,
\beq \label{Mgut} {g \vg}\simeq2 \cdot
10^{16}~\GeV\>\Rightarrow\>\vg\simeq2.86\cdot
10^{16}~\GeV~~\mbox{with}~~g\simeq0.7,\eeq
being the value of the unified gauge coupling constant. However,
we display in the following results for standard FHI which do not
fulfill \Eref{Mgut}. This is allowed since the relevant
restriction can be evaded if $G$ includes only abelian factors
(beyond the SM one) which do not disturb the gauge coupling
unification. Otherwise, threshold corrections may be taken into
account in order to restore the unification.

\paragraph{\bf (ii)} Boundness of $V_{\rm HI}$.
The inflationary potential is expected to be bounded from below.
This requirement lets open the possibility that the inflaton may
give rise to an inflationary expansion  under generic initial
conditions set at $\sigma\simeq\mP$.

\paragraph{\bf (iii)} Convergence of $V_{\rm HI}$. The expression
of $\Vhi$ in \Eref{Vol} is expected to converge at least for
$\sgm\sim\sgm_*$. This fact can be ensured if, for
$\sgm\sim\sgm_*$, each successive term in the expansion of $V_{\rm
SUGRA}$ (and $K$) \Eref{Vsugra1} (and \Eref{K}) is smaller than
the previous one. In practice, this objective can be easily
accomplished if the $k$'s in \Eref{K} -- or \Eref{Vol} -- are
sufficiently low.

\paragraph{\bf (iv)}  Monotonicity of $V_{\rm HI}$.
Depending on the values of the coefficients $k$'s in \Eref{Vol},
$V_{\rm HI}$ is a monotonic function of $\sigma$ or develops a
local minimum and maximum. The latter case leads to the possible
complication in which the system gets trapped near the minimum of
the inflationary potential and, consequently, no FHI takes place.
It is, therefore, crucial to check if we can avoid the
minimum-maximum structure of $V_{\rm HI}$. In such a case the
system can start its slow rolling from any point on the
inflationary path without the danger of getting trapped. This can
be achieved, if we require that $V_{\rm HI}$ is a monotonically
increasing function of $\sigma$, i.e. $V'_{\rm HI}>0$ for any
$\sigma$ or, equivalently,
\beq V'_{\rm HI}(\bar \sigma_{\rm min})>0~~\mbox{with}~~ V''_{\rm
HI}(\bar \sigma_{\rm min})=0~~\mbox{and}~~V'''_{\rm HI}(\bar
\sigma_{\rm min})>0\label{con}\eeq
where $\bar \sigma_{\rm min}$ is the value of $\sigma$ at which
the minimum of $V'_{\rm HI}$ lies.

\paragraph{\bf (v)} Tuning of the initial conditions. When hilltop FHI
occurs with $\sigma$ rolling from the region of the maximum down
to smaller values, a mild tuning of the initial conditions is
required \cite{gpp} in order to obtain acceptable $n_{\rm s}$'s.
In particular, the lower $n_{\rm s}$ we want to obtain, the closer
we must set $\sigma_*$ to $\sigma_{\rm max}$, where $\sigma_{\rm
max}$  is the value of $\sigma$ at which the maximum of $V_{\rm
HI}$ lies. To quantify somehow the amount of this tuning in the
initial conditions, we define \cite{gpp} the quantity:
\beq \Dex=\left(\sigma_{\rm max} - \sigma_*\right)/\sigma_{\rm
max}.\label{dms}\eeq
The naturalness of the attainment of FHI increases with $\Dex$.

\section{\scshape FHI in {\bf m}SUGRA}\label{msugra}

The simplest choice of \Ka emerging from the expression of
\Eref{K} is the one which assures canonical kinetic terms for the
inflaton field, $S$, with the minimal number of terms. This choice
is specified in \Sref{msugra1} and our results are discussed in
\Sref{msugra2}.

\subsection{\scshape The Relevant Set-up}\label{msugra1}

The used \Ka in this case can be derived from \Eref{K} by setting:
\beq
\K=0,~\Z=1~~\mbox{and}~~\ks=\kss=\ksss=\kst=\ksv=0.\label{mdef}\eeq
Upon substituting Eqs.~(\ref{c1k}) -- (\ref{c10k}) into \Eref{Vol}
we infer that the resulting $V_{\rm HI}$ takes the form
\beq\label{Vmsugra} V_{\rm HI}\simeq V_{\rm HI0}\left(1+c_{\rm
HI}-\aS{\sgm\over\sqrt{2\Vhio}}+{\sigma^4\over8\mP}\right),\eeq
since in this case $c^{(0)}_{2K}=0$ and $c^{(0)}_{4K}={1/2}$. It
is worth mentioning that mSUGRA is, in principle, beneficial for
the implementation of FHI, since it does not generate any new
contribution in the $\eta$ parameter, \Eref{slow}, due to a
miraculous cancellation emerging in the computation of
$c^{(0)}_{2K}$. Despite this fact, fixing all the remaining terms
in \Eref{K} beyond the quadratic term equal to zero can be
regarded as an ugly tuning.

\subsection{\scshape Results}\label{msugra2}

The investigation of this model of FHI depends on the parameters:
$$ \sigma_*, \vg, \aS~~\mbox{and}\left\{\bem
\kappa\hfill   & \mbox{for standard and shifted FHI}, \hfill \cr
M_{\rm S}\hfill  &\mbox{for smooth FHI}, \hfill \cr\eem
\right.$$
where we fix $M_{\rm S}=5\cdot 10^{17}~{\rm GeV}$ in the case of
shifted FHI. In our computation, we use as input parameters $\aS$
and $\kappa$ or $M_{\rm S}$. We then restrict $\vg$ and $\sigma_*$
so as Eqs.~(\ref{Nhi}) and (\ref{Prob}) are fulfilled. Using
Eqs.~(\ref{nS}) and (\ref{aS}) we can extract $n_{\rm s}$ and
$\alpha_{\rm s}$ respectively. Our findings for standard [shifted
and smooth] FHI are displayed in \Sref{stm} [\Sref{shm}].

We can obtain a rather accurate estimation of the expected $\ns$'s
if we omit the third term in the r.h.s of \Eref{Vmsugra},
calculate analytically the integral in Eq.~(\ref{Nefold}), replace
the $\sigma_{\rm f}$'s by their values in \Eref{sigmaf} and solve
the resulting equation w.r.t $\sigma_*$. Taking into account that
$\epsilon<\eta$ we can extract $\ns$ from Eq.~(\ref{nS}) and find
\begin{equation} \label{nssugra} n_{\rm s}\simeq\left\{\bem
1-{1/N_{\rm HI*}}+{3\kappa^2{\sf\ftn N}N_{\rm HI*}/4\pi^2} \hfill
& \mbox{for standard FHI}, \hfill \cr
1-{1/N_{\rm HI*}}+{3\kappa^2N_{\rm HI*}/2\pi^2} \hfill &\mbox{for
shifted FHI}, \hfill \cr
1-{5/3N_{\rm HI*}}+2\left(6\mu^2_{\rm S}M^2_{\rm S}N_{\rm
HI*}/m^4_{\rm P}\right)^{1/3}\hfill &\mbox{for smooth FHI}. \hfill
\cr\eem
\right.\end{equation}
Observing that the last term in the r.h.s of the expressions above
arise from the last term in the r.h.s of \Eref{Vmsugra}, we can
easily infer that mSUGRA increases significantly $n_{\rm s}$ for
relatively large $\kp$'s or $M_{\rm S}$'s.

\begin{figure}[!t]\vspace*{-.16in}
\hspace*{-.2in}
\begin{minipage}{8in}
\epsfig{file=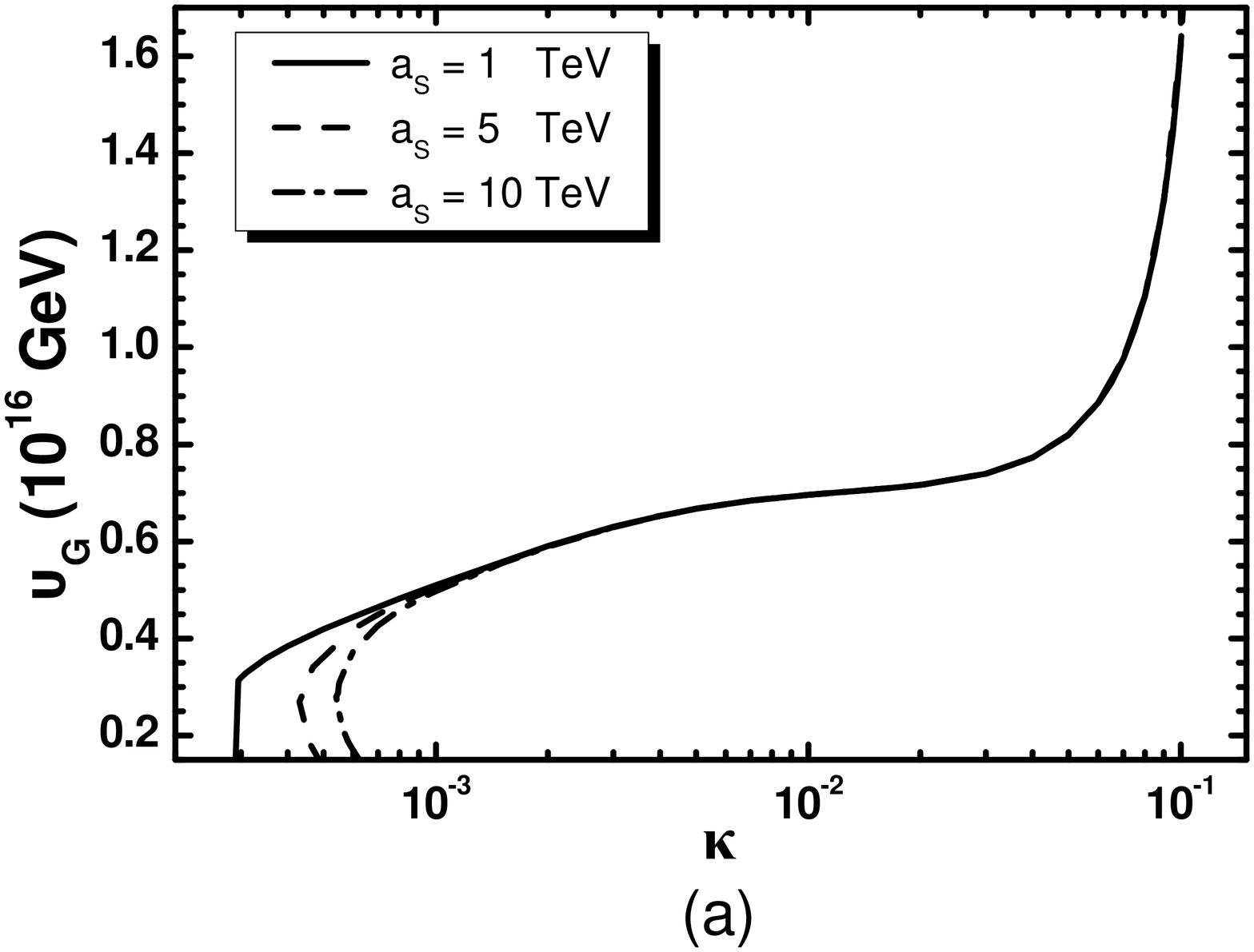,height=3.25in,angle=-90}
\hspace*{-1.25cm}
\epsfig{file=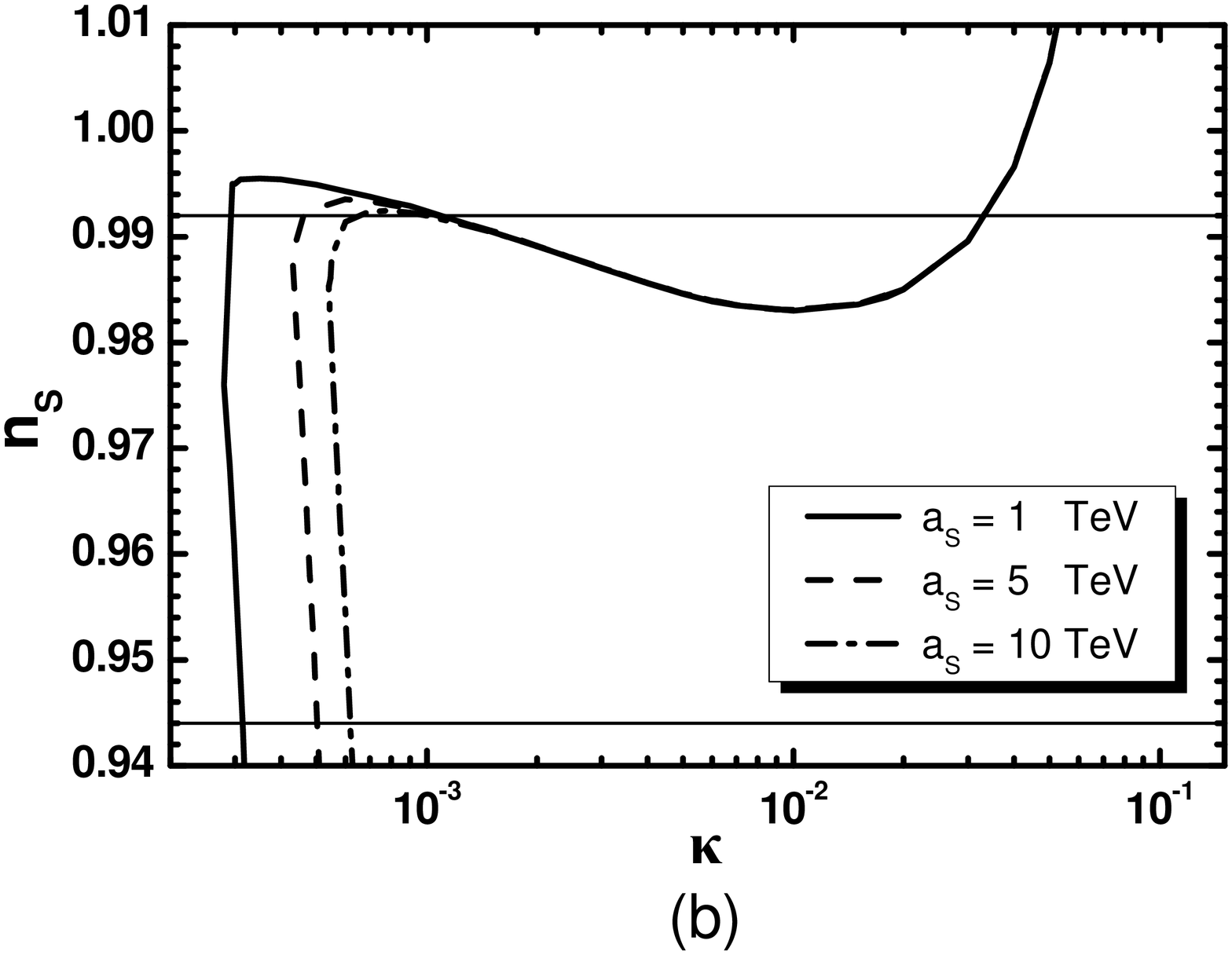,height=3.25in,angle=-90} \hfill
\end{minipage}
\hfill \caption[]{\ftn\sl The allowed by Eqs.~(\ref{Nhi}) and
(\ref{Prob}) values of $\vg$ [$n_{\rm s}$] versus $\kappa$ for
standard FHI with ${\sf\ftn N}=2$ and $\aS=1~{\rm TeV}$ (solid
lines), $\aS=5~{\rm TeV}$ (dashed lines) and $\aS=10~{\rm TeV}$
(dot-dashed lines). The region of Eq.~(\ref{nswmap}) is also
limited by thin lines.}\label{msugraf}
\end{figure}

On the other hand, the third term in the r.h.s of \Eref{Vmsugra}
can be important for $\kappa\leq6\cdot10^{-4}$ and
$M\leq10^{15}~\GeV$ -- cf. \cref{sstad}-- since it becomes
comparable with the second term there. In this regime, the
required by \Eref{Nhi} $\sgm_*$ becomes comparable to $\sgm_{\rm
c}$ and the approximation of $f_{\rm rc}$ in \Eref{frc1} is no
longer valid. Instead, we here have
\beq f_{\rm rc}(x)=3 - {x^{-2}\over6} - {x^{-4}\over30} -
{x^{-6}\over84} - {x^{-8}\over180} - {x^{-10}\over330} -
{x^{-12}\over546} - {x^{-14}\over840} - {x^{-16}\over1224} -
{x^{-18}\over1710} - {x^{-20}\over2310}-\cdots\label{frc2}\eeq
As a consequence, $\Vhi'$ decreases sharply (enhancing $\Nhi$)
whereas $|\Vhi''|$ (or $\eta$) increases adequately, lowering
thereby $\ns$ to an acceptable level.

\subsubsection{\scshape Standard FHI} \label{stm}

In the case of standard FHI (with ${\sf\ftn N}=2$), we display in
\sFref{msugraf}{a} the allowed by Eqs.~(\ref{Nhi}) and
(\ref{Prob}) values of $\vg$ versus $\kappa$. The corresponding
variation of $n_{\rm s}$ versus $\kappa$ is depicted in
\sFref{msugraf}{b} where the observationally compatible region of
Eq.~(\ref{nswmap}) is also delimited by thin lines. Solid, dashed
and dot-dashed lines stand for the results obtained for $\aS=1,5$
and $10~{\rm TeV}$ respectively. We observe that the various lines
coincide for $\kappa\gtrsim6\cdot10^{-4}$. For the sake of clarity
we do not show in \Fref{msugraf} solutions with
$\vg>2\cdot10^{16}~\GeV$ or $\kp<2\cdot10^{-4}$ -- cf.
\cref{sstad} -- which are totally excluded by \Eref{nswmap}. The
third [last] term in the r.h.s of \Eref{Vmsugra} become important
for $\kappa\gtrsim0.01$ [$\kappa\lesssim6\cdot10^{-4}$] whereas
for $6\cdot10^{-4}\lesssim\kappa\lesssim0.01$, the second term in
the r.h.s of \Eref{Vmsugra} becomes prominent. As a consequence,
the last term in the r.h.s of \Eref{Vmsugra} drives $n_{\rm s}$ to
values close to or larger than unity whereas the third one succeed
in reconciling it with \Eref{nswmap} for discriminated $\kappa$'s
related to the chosen $\aS$'s. Namely, from \Fref{msugraf} we
deduce that there is a marginally allowed area for
\beqs\bea \label{rem1} &&
0.0015\lesssim\kappa\lesssim0.032,~~5.5\lesssim
\vg/(10^{15}~{\rm GeV}) \lesssim7.5,\\
&& 0.983\lesssim n_{\rm s}\lesssim0.99~~\mbox{and}~~1.2\lesssim
|\as|/10^{-4} \lesssim3.5. \label{rem2}\eea
In addition, from \Fref{msugraf} we find isolated corridors
consistent with \Eref{nswmap}, e.g.
\beq
\label{rem3}\kappa\simeq3\cdot10^{-4},~5\cdot10^{-4},~6\cdot10^{-4}
~~\mbox{with}~~\vg\lesssim3\cdot10^{15}~\GeV~~\mbox{for}~~\aS=1,
5~~\mbox{and}~~10~\TeV,\eeq\eeqs
respectively. We remark that the $\vg$'s allowed here lie well
below the ones required by \Eref{Mgut}. In conclusion, although
standard FHI in mSUGRA can not be excluded, it can be considered
as rather disfavored since the allowed region is extremely
limited.

\begin{table}[!t]
\begin{center}
\begin{tabular}{|l|l|l||l|l|}
\hline
\multicolumn{3}{|c||}{\sc Shifted FHI}&\multicolumn{2}{|c|}{\sc
Smooth FHI}\\ \hline \hline
$\kappa/10^{-3}$ &\multicolumn{2}{|c||}{$9.2$}&$M_{\rm
S}/5\cdot10^{17}~{\rm GeV}$ &$0.79$\\
$\sigma_*/10^{16}~{\rm GeV}$
&\multicolumn{2}{|c||}{$5.37$}&$\sigma_*/10^{16}~{\rm GeV}$ &
$32.9$\\ \hline\hline
$M/10^{16}~{\rm GeV}$&\multicolumn{2}{|c||}{$2.3$}&$\mu_{\rm
S}/10^{16}~{\rm GeV}$&$0.21$
\\
$1/\xi$ &\multicolumn{2}{|c||}{$4.36$}&$\sigma_{\rm
f}/10^{16}~{\rm GeV}$&$13.4$
\\
$N_{\rm HI*}$ &\multicolumn{2}{|c||}{$52.2$}&$N_{\rm HI*}$
&$53$\\
$n_{\rm s}$ &\multicolumn{2}{|c||}{$0.982$}&$n_{\rm s}$ &
$1.04$\\
$-\alpha_{\rm s}/10^{-4}$ & \multicolumn{2}{|c||}{$3.4$}&
$-\alpha_{\rm s}/10^{-4}$ &$16.6$\\ \hline
\end{tabular}
\end{center}
\caption{\sl\ftn Input and output parameters consistent with
Eqs.~(\ref{Nhi}), (\ref{Prob}) and (\ref{Mgut}) for shifted (with
$M_{\rm S}=5\cdot10^{17}~{\rm GeV}$) or smooth FHI within
mSUGRA.}\label{tabhi}
\end{table}

\subsubsection{\scshape Shifted and Smooth FHI}\label{shm}

In the cases of shifted and smooth FHI we confine ourselves to the
values of the parameters consistent with \Eref{Mgut} and display
the solutions fulfilling Eqs.~(\ref{Nhi}) and (\ref{Prob}) in
Table~\ref{tabhi}. Given that $\aS\sim1~\TeV$ plays no role in the
determination of the inflationary observables, we conclude that
the resulting $\ns$'s and $\as$'s are obviously predictions of
these FHI models -- without the possibility of altering them by
some adjustment. We observe that the required $\kappa$, in the
case of shifted FHI, is rather low and so, the last terms in the
r.h.s of \Eref{Vmsugra} is more or less negligible. As a result,
$\eta$ is exclusively determined  by $\chir\Vhio$ and $n_{\rm s}$
remains within the range of Eq.~(\ref{nswmap}) although outside
the $68\%$ c.l. region. On the contrary, in the case of smooth
FHI, $\eta$ strongly depends on the last term in the r.h.s of
\Eref{Vmsugra} enhancing thereby $n_{\rm s}$ beyond the range of
\Eref{nswmap}. In the latter case, $|\alpha_{\rm s}|$ is also
considerably enhanced.

\section{\scshape Hilltop FHI in {\bf nm}SUGRA}\label{nmsugra}

The fitting of the WMAP7 data with the $\Lambda$CDM model enforces
\cite{gpp, mur} us to consider more complicated (and possibly more
general) forms of K\"alher potentials. The simplest choice is to
consider of a moderate deviation from mSUGRA, named \cite{gpp}
nmSUGRA, according to which the next-to-minimal term is tuned so
as an adequately small, negative mass squared for the inflaton is
generated. This setting is outlined in \Sref{nmsugra1} and the
structure of the resulting $V_{\rm HI}$ is analyzed in
Sec.~\ref{nmsugra2}. Our results are exhibited in
Sec.~\ref{nmsugra3}.

\subsection{\scshape The Relevant Set-up}\label{nmsugra1}

In this scenario, the form of the relevant K\"{a}hler potential is
given by \Eref{K}, setting
\beqs\beq \K=\kss=\ksss=\kst=\ksv=0~~\mbox{and}~~
\Z=1.\label{nmdef}\eeq
Therefore, $V_{\rm HI}$ takes the form of \Eref{Vol} with
\beq\label{cnm} c_{2K}=c^{(0)}_{2K}=\ks,~
c_{4K}=c^{(0)}_{4K}(\kss=0)~~\mbox{and}~~c_{2\nu
K}=0~~\mbox{for}~~\nu\geq3.\eeq\eeqs
Note that the expansion of $V_{\rm SUGRA}$ in \Eref{Vsugra1}
terminates at the terms of order $\p^4$ consistently with the fact
that the expansion of $K$ terminates at the terms of order $\p^6$.

\subsection{\scshape Structure of the Inflationary Potential}\label{nmsugra2}

For $\sgm$ close to $\sgm_*$, $\Vhi$ given by \Eref{Vol} can be
approximated as
\beq\label{Vnm} \Vhi\simeq\Vhio\,\left(1+\ c_{\rm
HI}-\,\ks{\sigma^2\over2\mP^2}+\,\ckk{\sigma^4\over4\mP^4}\right)\cdot\eeq
Given that $\ckk=\ckk^{(0)}\simeq1/2$ is much larger than $\ks$,
the boundedness of $\Vhi$ is ensured in this scenario, by
construction.

The monotonicity of $\Vhi$ in \Eref{Vnm} can be investigated
applying Eq.~(\ref{con}). In particular, we can find approximately
-- note that in the formulas below, we have $\ckk\simeq1/2$:
\beq \label{smin1} \bar\sigma_{\rm min}\simeq\left\{\bem
\sqrt{\ks/3\ckk}\; m_{\rm P}\hfill  &\mbox{for standard and
shifted FHI}, \hfill \cr
\sqrt{2m_{\rm P}/3}\left(\sqrt{5/2\ckk}\;\mu_{\rm S}M_{\rm
S}\right)^{1/4}\hfill &\mbox{for smooth FHI}. \hfill \cr\eem
\right.\end{equation}
Inserting Eq.~(\ref{smin1}) into Eq.~(\ref{con}), we find that
$V_{\rm HI}$ remains monotonic for
\beq \ks<\ks^{\rm max}~~\mbox{with}~~\ks^{\rm max}=\left\{\bem
{3\kappa\sqrt{\ckk{\sf\ftn N}}/4\pi}\hfill   & \mbox{for standard
FHI}, \hfill \cr
{3\kappa\sqrt{\ckk}/2\sqrt{2}\pi}\hfill  &\mbox{for shifted FHI},
\hfill \cr
(8/3)(2\ckk/5)^{3/4}\sqrt{\mu_{\rm S}M_{\rm S}}/m_{\rm P}\hfill
&\mbox{for smooth FHI}. \hfill \cr\eem
\right. \label{cqmax}\eeq

\begin{figure}[!t]\vspace*{-.3in}
\begin{center}
\epsfig{file=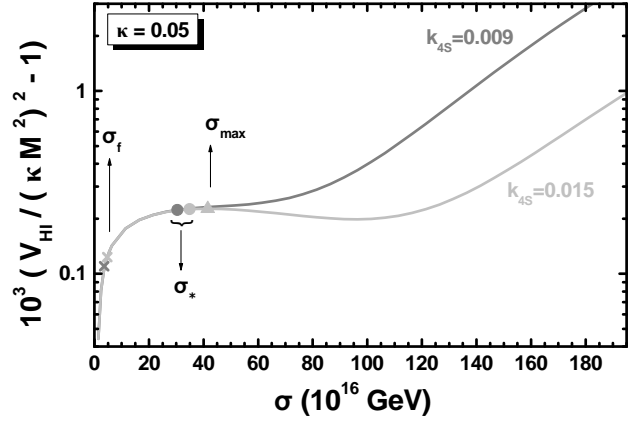,height=3.65in,angle=-90}
\end{center}
\hfill \caption[]{\sl\ftn The variation of $\Vhi$ in \Eref{Vnm} as
a function of $\sgm$ for standard FHI in nmSUGRA, $\kappa=0.05$
and $\ks=0.009$ ($\ns=0.968$) [$\ks=0.015$ ($\ns=0.944$)](gray
[light gray] line). The values of $\sigma_*, \sigma_{\rm f}$ and
$\sgm_{\rm max}$ are also depicted.} \label{Vhiq}
\end{figure}

For $\ks>\ks^{\rm max}$, $V_{\rm HI}$ reaches a local minimum
[maximum] at the inflaton-field value $\sigma_{\rm min}$
[$\sigma_{\rm max}$] which can be estimated as follows:
\begin{equation}
\label{sigmamax} \sigma_{\rm min}\simeq \sqrt{\ks\over\ckk} m_{\rm
P} ~~\mbox{and}~~\sigma_{\rm max}\simeq\left\{\bem
\kappa m_{\rm P} \sqrt{\sf\ftn N}/2\sqrt{2\ks}\pi\hfill &
\mbox{for standard FHI}, \hfill \cr
{\kappa m_{\rm P}/2\sqrt{\ks}\pi}\hfill &\mbox{for shifted FHI},
\hfill \cr
\sqrt{2/3\sqrt[3]{\ks}}{\left(\mu_{\rm S}M_{\rm S}m_{\rm
P}\right)^{1/3}}\hfill &\mbox{for smooth FHI}. \hfill \cr\eem
\right.\end{equation}

The structure of $\Vhi$ is depicted in \Fref{Vhiq} where we
display the variation of $\Vhi$ as a function of $\sgm$ for
standard FHI in nmSUGRA, $\kappa=0.05$ and $\ks=0.009$ (gray line)
or $\ks=0.015$ (light gray line). In the first case (gray line) we
obtain $\ns=0.968$ with $\ks<\ks^{\rm max}\simeq0.011$ and
therefore $\Vhi$ remains monotonic. On the contrary, for
$\ks=0.015$ we get $\ns=0.944$ with $\ks>\ks^{\rm max}$ and
therefore $\Vhi$ develops the minimum-maximum structure with the
maximum being located at $\sgm_{\rm max}=4.15\cdot10^{17}~\GeV$.
The resulting $\Dex$ in the latter case is $\Dex=0.16$. The values
of $\sigma_*$ and $\sigma_{\rm f}$ are also depicted.

\subsection{\scshape Results}\label{nmsugra3}

Our strategy in the numerical investigation of the nmSUGRA
scenario is the one described in Sec.~\ref{msugra2} -- recall that
we fix $\aS=1~\TeV$ henceforth and no impact from that term on our
results is detected. In addition to the parameters manipulated
there, we have here the parameter $\ks$ which can be adjusted so
as to achieve $n_{\rm s}$ in the range of Eq.~(\ref{nswmap}). We
check also the fulfillment of Eq.~(\ref{con}). Our findings for
standard [shifted and smooth] FHI are accommodated in \Sref{stnm}
[\Sref{shnm}].

Employing the procedure outlined in Sec.~\ref{msugra2} above
\Eref{nssugra} we can take a flavor for the expected $n_{\rm s}$'s
in the nmSUGRA scenario, for any $\ks$:
\begin{equation} \label{nsq} n_{\rm s}\simeq\left\{\bem
1-2\ks\left(1-1/c_N\right)-{6\ckk\kappa^2{\sf\ftn N}
c_N/4\ks\pi^2}\hfill & \mbox{for standard FHI}, \hfill \cr
1-2\ks\left(1-1/c_N\right)-{6\ckk\kappa^2 c_N/2\ks\pi^2}\hfill
&\mbox{for shifted FHI}, \hfill \cr
1-5/3N_{\rm HI*}+2\tilde c_N-\left(2\tilde c_N N_{\rm
HI*}+7\right)\ks\hfill &\mbox{for smooth FHI}, \hfill \cr\eem
\right.\end{equation} $$\mbox{with}~~ c_N=1-\sqrt{1+4\ks N_{\rm
HI*}}~~\mbox{and}~~ \tilde c_N=2\ckk\left(6\mu^2_{\rm S}M^2_{\rm
S}N_{\rm HI*}/m^4_{\rm P}\right)^{1/3}.$$
We can clearly appreciate the contribution of a positive $\ks$ in
lowering of $n_{\rm s}$.

\subsubsection{\scshape Standard FHI} \label{stnm}

\begin{figure}[!t]\vspace*{-.16in}
\hspace*{-.2in}
\begin{minipage}{8in}
\epsfig{file=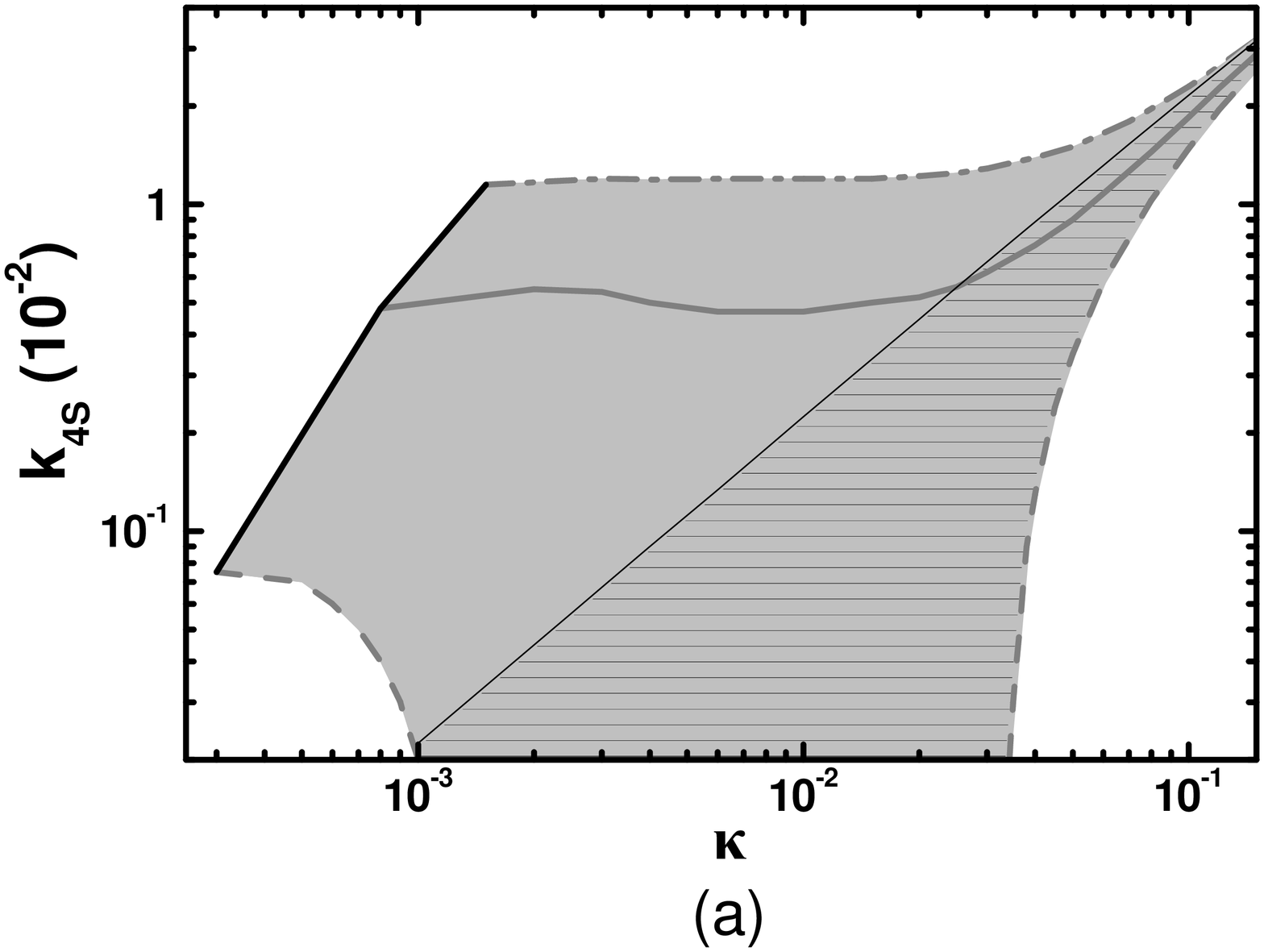,height=3.25in,angle=-90}
\hspace*{-1.25cm}
\epsfig{file=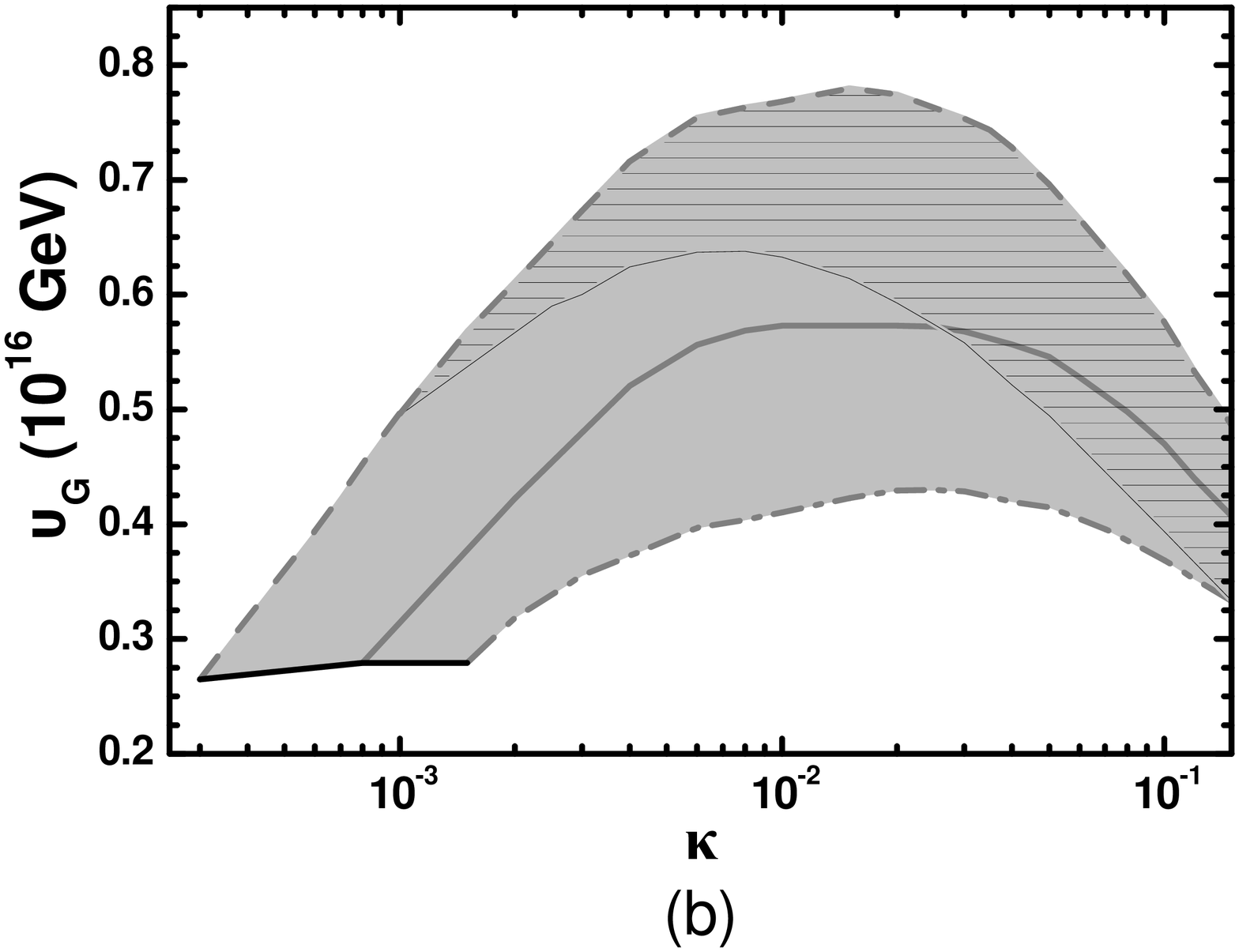,height=3.25in,angle=-90} \hfill
\end{minipage}\vspace*{-.1in}
\begin{center}
\epsfig{file=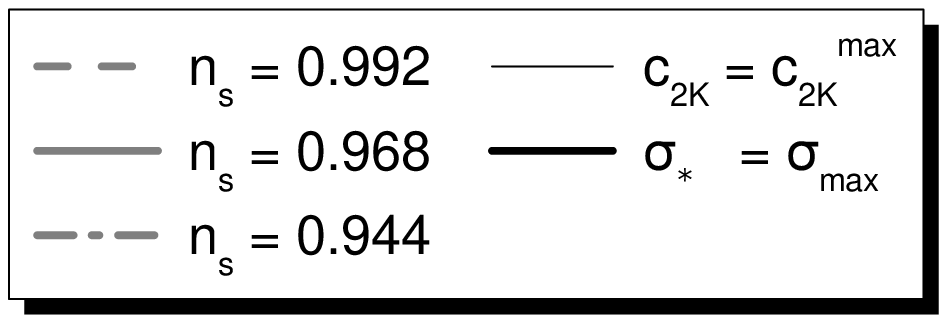,height=1.8in,angle=-90}
\end{center}
\hfill \caption[]{\sl\ftn Allowed (lightly gray shaded) region as
determined by Eqs.~(\ref{Nhi})-(\ref{aswmap}) in the
$\kappa-c_{2K}$ [$\kappa-v_{_G}$] plane (a) [(b)] for standard FHI
within the nmSUGRA. Ruled are the regions where the inflationary
potential remains monotonic. The conventions adopted for the
various lines are also shown.}\label{nmf}
\end{figure}

In the case of standard FHI (with ${\sf\ftn N}=2$), we delineate
the (lightly gray shaded) region allowed by
Eqs.~(\ref{Nhi})-(\ref{aswmap}) in the $\kappa-\ks$ [$\kappa-\vg$]
plane -- see \sFref{nmf}{a} [\sFref{nmf}{b}]. The conventions
adopted for the various lines are also shown. In particular, the
dashed [dot-dashed] lines correspond to $n_{\rm s}=0.992$ [$n_{\rm
s}=0.944$], whereas the gray solid lines are obtained by fixing
$n_{\rm s}=0.968$ -- see Eq.~(\ref{nswmap}). Below the black solid
lines, our initial assumption $\sigma_*<\sigma_{\rm max}$ is
violated. In the hatched regions, Eq.~(\ref{con}) is also
satisfied and along their boundaries designed by thin, black,
solid lines Eq.~(\ref{cqmax}) is saturated. We observe that the
optimistic constraint of Eq.~(\ref{con}) can be met in a rather
wide fraction of the allowed area. In particular, for $n_{\rm
s}=0.968$ we find
\beqs\bea \label{renm1} &&
\{0.8\}~2.5\lesssim{\kp\over10^{-2}}\lesssim15,~~\{11\}~
5.7\gtrsim {\vg\over10^{15}~{\rm GeV}}\gtrsim4.1, \\ \label{renm2}
&& \{0.48\} ~0.56\lesssim
{\ks\over10^{-2}}\lesssim2.9~~\mbox{and}~~\{0.2\}~ 0.39\lesssim
-{\as\over10^{-3}}\lesssim1.1,\eea\eeqs
where the limiting values obtained without imposing \Eref{con} are
indicated in curly brackets. In the corresponding region, $\Dex$
ranges between $0$ and $50\%$. Note that the $\vg$'s encountered
here are lower that those required by \Eref{Mgut}.

\subsubsection{\scshape Shifted and Smooth FHI}\label{shnm}

In the cases of shifted and smooth FHI we confine ourselves to the
values of the parameters which satisfy \Eref{Mgut} and display in
Table~\ref{tabq} their values which are consistent with
Eqs.~(\ref{Nhi})-(\ref{aswmap}) as well. In the case of shifted
FHI, we observe that (i) we need positive $\ks$ to obtain $n_{\rm
s}=0.992$ since the mSUGRA result is lower -- see
Table~\ref{tabhi}; (ii) the lowest possible $n_{\rm s}$ compatible
with the conditions of Eq.~(\ref{con}) is $0.976$ and so, $n_{\rm
s}=0.968$ is not consistent with Eq.~(\ref{con}). In the case of
smooth FHI, we see that a reduction of $n_{\rm s}$ consistently
with Eq.~(\ref{con}) can be achieved for $n_{\rm s}\gtrsim 0.951$
and so $n_{\rm s}=0.968$ can be obtained without complications.

\begin{table}[!t]
\begin{center}
\begin{tabular}{|l|lll||l|lll|}
\hline
\multicolumn{4}{|c||}{\sc Shifted FHI}&\multicolumn{4}{|c|}{\sc
Smooth FHI}\\ \hline\hline
$n_{\rm s}$ &  $0.944$&$0.968$&$0.992$&$n_{\rm s}$ &
$0.944$&$0.968$&$0.992$\\
$\ks/10^{-3}$ &  $11.7$&$4.5$&$-3.2$&$\ks/10^{-3}$ &
$9.65$&$7.6$&$5.4$\\
$\ks^{\rm max}/10^{-3}$ &  $1.8$&$1.9$&$2.2$&$\ks^{\rm
max}/10^{-3}$ & $9.1$&$9.2$&$9.2$\\\hline
$\Delta_{\rm m*}/10^{-2}$ &  $16$&$39$&$-$&$\Delta_{\rm
m*}/10^{-2}$ & $7$&$-$&$-$\\\hline\hline
$\sigma_*/10^{16}~{\rm GeV}$ &$2.45$
&$3.1$&$4.31$&$\sigma_*/10^{16}~{\rm GeV}$ &
$23.8$&$24.9$&$26.6$\\
$\kappa/10^{-3}$ & $8.1$&$8.7$&$9.7$&$M_{\rm S}/5\cdot
10^{17}~{\rm GeV}$ & $2.4$&$1.83$&$1.4$\\ \hline
$M/10^{16}~{\rm GeV}$&$2.2$& $2.26$&$2.3$&$\mu_{\rm
S}/10^{16}~{\rm GeV}$& $0.07$&$0.09$&$0.12$\\
$1/\xi$ &  $4.1$&$4.3$&$4.5$&$\sigma_{\rm f}/10^{16}~{\rm
GeV}$&$13.4$&$13.4$&$13.4$ \\
$N_{\rm HI*}$ &  $51.9$&$52.1$&$52.3$&$N_{\rm HI*}$ &
$52.3$&$52.4$&$52.6$\\
$-\alpha_{\rm s}/10^{-4}$ &  $3.1$&$3.5$&$3.5$ &$-\alpha_{\rm
s}/10^{-3}$ & $0.7$&$0.9$&$1.1$\\ \hline
\end{tabular}
\end{center}
\caption{\sl\ftn Input and output parameters consistent with
Eqs.~(\ref{Nhi}) - (\ref{Mgut}) for shifted (with $M_{\rm
S}=5\cdot 10^{17}~{\rm GeV}$) or smooth FHI in
nmSUGRA.}\label{tabq}
\end{table}

\section{\scshape Hilltop FHI in {\bf nnm}SUGRA}\label{nnmsugra}

Another possible SUGRA set-up which can accommodate an
observationally viable FHI is the one, first proposed in
\cref{rlarge}, which we here name nnmSUGRA. In this case, a
convenient choice, specified in \Sref{nnmsugra1}, of the
next-to-minimal and the next-to-next-to-minimal terms in the \Ka
is employed. The structure of the resulting $V_{\rm HI}$ is
studied in Sec.~\ref{nnmsugra2} and our related results are
exhibited in Sec.~\ref{nnmsugra2}.

\subsection{\scshape The Relevant Set-up}\label{nnmsugra1}

In this scenario, the form of the relevant K\"{a}hler potential is
given by \Eref{K} after setting 
\beqs\beq \K=0~~\mbox{and}~~\Z=1\label{nnmdef} \eeq
and so $V_{\rm HI}$ takes the form of \Eref{Vol} with
\beq\label{cnnm}
c_{2K}=c^{(0)}_{2K}=\ks,~~c_{4K}=c^{(0)}_{4K},~~c_{6K}=c^{(0)}_{6K},~~c_{8K}=c^{(0)}_{8K}
~~\mbox{and}~~c_{10K}=c^{(0)}_{10K}.\eeq\eeqs
In other words, this scenario is the most general one which arises
from the K\"{a}hler potential of \Eref{K} in the absence of
$h_m$'s. The crucial difference between nnmSUGRA and nmSUGRA,
however is the sign of $\ck=\ks$ which is here negative. As a
consequence, fulfilling of \Eref{nswmap} requires negative $\ckk$
or positive $\kss$ -- see \Eref{c6k}. The inclusion of higher
order terms in the expansion of \Eref{Vsugra1} prevents the
runaway behavior of the resulting $\Vhi$ -- see \Eref{Vol}.

\subsection{\scshape Structure of the Inflationary Potential}\label{nnmsugra2}

For $\sgm$ close to $\sgm_*$, $\Vhi$ given by \Eref{Vol} can be
approximated as -- cf. \cref{nolde}:
\beq\label{Vnnm} \Vhi\simeq\Vhio\,\left(1+\ c_{\rm
HI}-\,\ks{\sigma^2\over2\mP^2}+\,\ckk{\sigma^4\over4\mP^4}-\,\ckx{\sigma^6\over8\mP^6}\right)\cdot\eeq
The monotonicity of $\Vhi$ here can be checked only numerically --
due to the numerous terms involved in $\Vhi'$ and $\Vhi''$ -- by
applying the criterion in \Eref{con}. In the case of a
non-monotonic $\Vhi$, we can show that it reaches a local maximum
at the inflaton-field value:
\beq \label{sigmamax2}\sigma_{\rm max}\simeq\left\{\bem
{\mP \sqrt{2\pi|\ks| + \sqrt{2}\sqrt{2\ks^2\pi^2 + {\sf\ftn N}k^2
|\ckk|}}\over2 \sqrt{\pi|\ckk|}} \hfill & \mbox{for standard FHI},
\hfill \cr
{\mP \sqrt{\pi|\ks| + \sqrt{\ks^2\pi^2 + k^2|\ckk|}}\over
\sqrt{2\pi|\ckk|}}\hfill &\mbox{for shifted FHI}, \hfill \cr
(2/3)^{3/8} \sqrt{\mP\sqrt{\mu_{\rm S}M_{\rm S}}} /|\ckh|^{1/8}
\hfill &\mbox{for smooth FHI}. \hfill \cr\eem
\right.\end{equation} and a local minimum at the inflaton-field
value:
\begin{equation}
\label{sigmamin2} \sigma_{\rm min}\simeq \mP{\sqrt{-3 |\ckx| +
\sqrt{9 \ckx^2 + 32 |\ckk \ckx|}}\over2 \sqrt{|\ckh|}}\cdot\eeq
The last result holds for all the types of FHI, since for $\sgm$'s
close $\sigma_{\rm min}$, $\Vhi$ is dominated by the last terms of
the expansion in the r.h.s of \Eref{Vol} and so, any depedence on
$c_{\rm HI}$, which essentially indentifies the type of FHI, is
switched off.

\begin{figure}[!t]\vspace*{-.3in}
\begin{center}
\epsfig{file=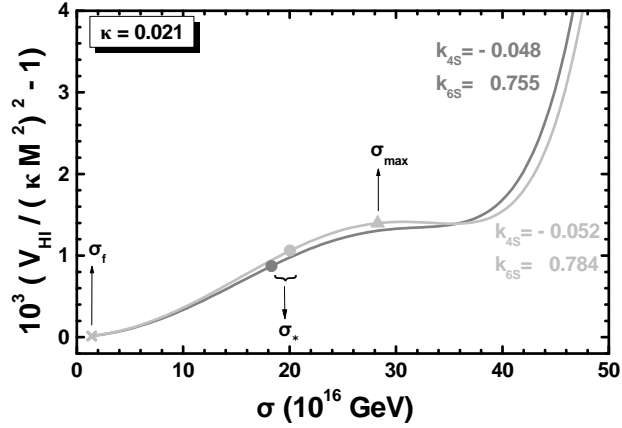,height=3.65in,angle=-90}
\end{center}
\hfill \caption[]{\sl\ftn The variation of $\Vhi$ in \Eref{Vnnm}
as a function of $\sgm$ for standard FHI in nnmSUGRA,
$\kappa=0.021$, $\ksss=-1.5,\kst=-1,\ksv=0.5$ and
$\ks=-0.048,\kss=0.755$ ($\ns=0.968$) [$\ks=-0.052,\kss=0.784$
($\ns=0.944$)] (gray [light gray] line). The values of $\sigma_*,
\sigma_{\rm f}$ and $\sgm_{\rm max}$ are also depicted.}
\label{Vnnmf}
\end{figure}

The structure of $\Vhi$ is depicted in \Fref{Vnnmf} where we
display the variation of $\Vhi$ as a function of $\sgm$ for
standard FHI in nnmSUGRA, $\kappa=0.021,
\ksss=-1.5,\kst=-1,\ksv=0.5$ and $\ks=-0.048,\kss=0.755$ (gray
line) or $\ks=-0.052,\kss=0.784$ (light gray line). In the first
case (gray line) we obtain $\ns=0.968$ and $\Vhi$ remains
monotonic. On the contrary, in the second case (light gray line)
$\Vhi$ develops the minimum-maximum structure and we get
$\ns=0.944$. The maximum of $\Vhi$ is located at $\sgm_{\rm
max}=2.8\cdot10^{17}~\GeV$ and we get $\Dex=0.29$. The values of
$\sigma_*$ and $\sigma_{\rm f}$ are also depicted.

\subsection{\scshape Results}\label{nnmsugra3}

Our strategy in the numerical investigation of the nnmSUGRA
scenario is the one described in Sec.~\ref{msugra2}. In addition
to the parameters manipulated there, we have here the parameters
$\ks$ and $\kss$ which can be adjusted in order to fulfill
Eq.~(\ref{nswmap}) whereas the boundedness of $\Vhi$ is controlled
by the $k_{2\nu S}$'s with $4\leq\nu\leq6$. We check also the
validity of Eq.~(\ref{con}). Our findings for standard [shifted
and smooth] FHI are arranged in \Sref{stnnm} [\Sref{shnnm}].

Preliminarily results on the $\ns$'s expected, however, can be
extracted by applying the procedure highlighted above
\Eref{nssugra}. Namely, for standard FHI we find
\beqs\beq \ns \simeq   1 - {1\over2} \lf2|\ks| + {1\over\pi} \lf3
\sqrt{4\Delta_{1K}} \tanh{N_{\rm HI*}\sqrt{\Delta_{1K}}\over\pi} +
{\rm arctanh}{\Delta_{2K}\pi\over\sqrt{\Delta_{1K}}}\rg\rg\eeq
where \beq \label{Dnnm} \Delta_{1K} = (|\ckk| k^2 {\sf\ftn N} + 2
\ks^2 \pi^2)/2~~\mbox{and}~~\Delta_{2K} =-4|\ckk| M^2/\mP^2+|\ks|.
\eeq\eeqs The result for shifted FHI can be obtained from the
expression above setting ${\sf\ftn N}=2$ and replacing $4$ with
$2$ in the formula of $\Delta_{2K}$ in \Eref{Dnnm}. We did not
succeed to obtain similar formulas for smooth FHI due to
complications related to the numerus significant terms in $\Vhi$,
\Eref{Vnnm}.

\begin{figure}[!t]\vspace*{-.16in}
\hspace*{-.2in}
\begin{minipage}{8in}
\epsfig{file=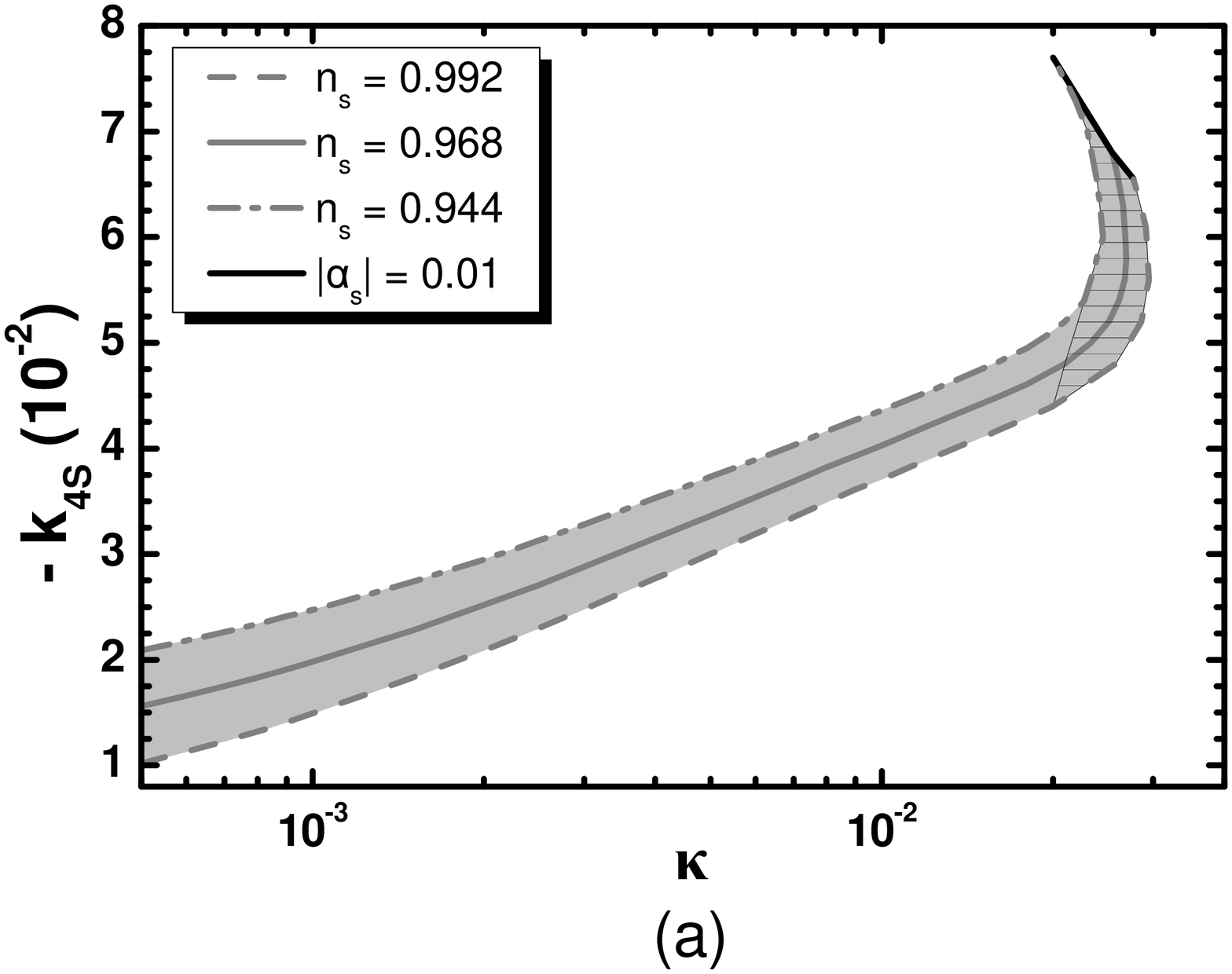,height=3.25in,angle=-90}
\hspace*{-1.25cm}
\epsfig{file=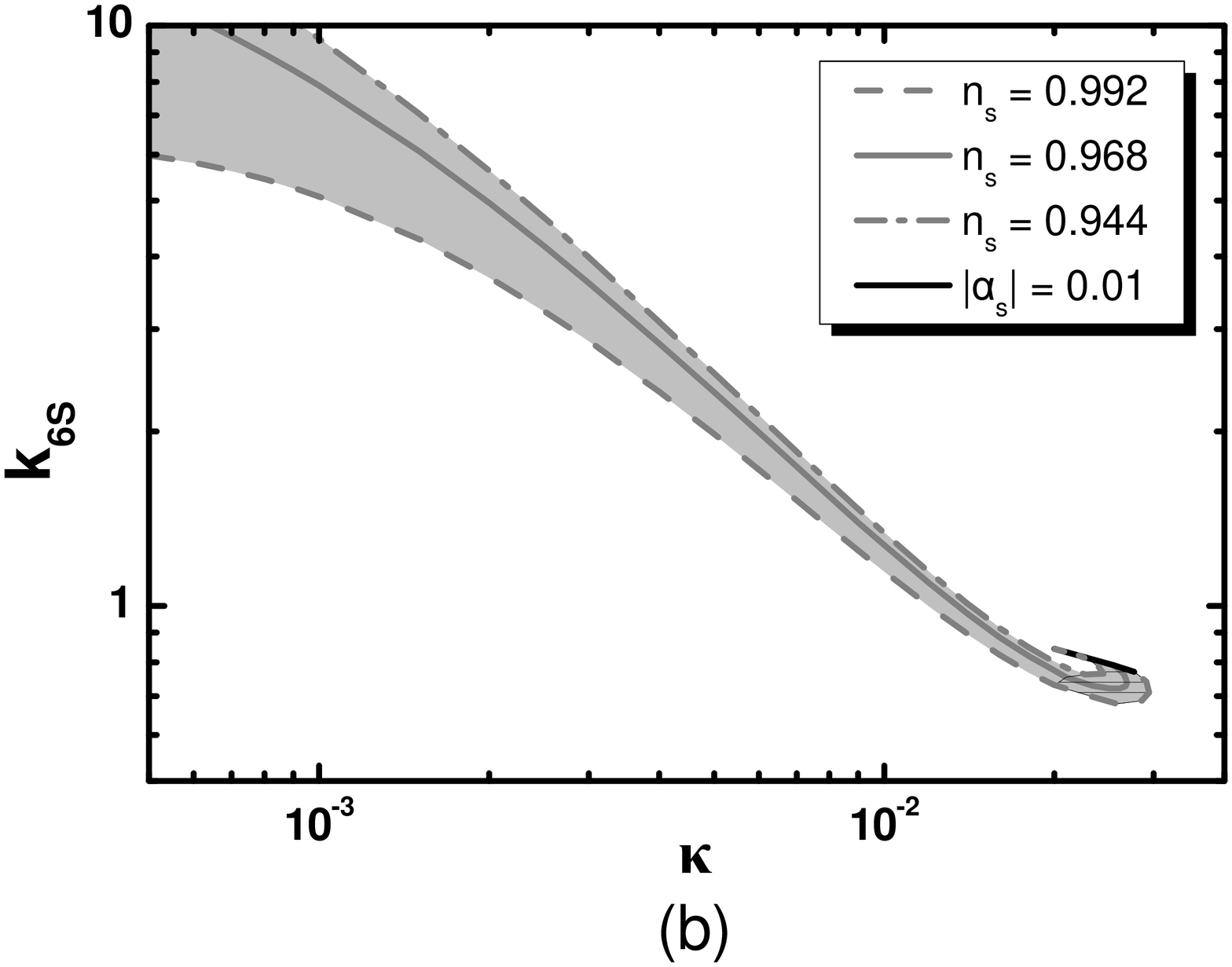,height=3.25in,angle=-90} \hfill
\end{minipage}
\hfill \caption[]{\sl\ftn Allowed (lightly gray shaded) region, as
determined by Eqs.~(\ref{Nhi})-(\ref{Mgut}), in the
$\kappa-(-\ks)$ [$\kappa-\kss$] plane (a) [(b)] for standard FHI
in nnmSUGRA. We take $\ksss=-1.5,\kst=-1$ and $\ksv=0.5$. Hatched
are the regions where $\Vhi$ remains monotonic. The conventions
adopted for the various lines are also shown.}\label{nnmf}
\end{figure}

\subsubsection{\scshape Standard FHI}\label{stnnm}

One of the outstanding advantages of the realization of standard
FHI within nnmSUGRA is that \Eref{Mgut} can be attained -- cf.
Figs. \ref{msugraf}-{\sf\ftn (a)}, \ref{nmf}-{\sf\ftn (b)} and
\ref{fig2}-{\sf\ftn (b)}, below. Therefore, in this scenario, we
are able to display regions (lightly gray shaded) allowed by
Eqs.~(\ref{Nhi}) -- (\ref{Mgut}) in the $\kappa-(-\ks)$
[$\kappa-\kss$] plane -- see \sFref{nnmf}{a} [\sFref{nnmf}{b}] --
for $\ksss=-1.5,\kst=-1, \ksv=0.5$. The conventions adopted for
the various lines are also shown. In particular, the gray dashed
[dot-dashed] lines correspond to $n_{\rm s}=0.992$ [$n_{\rm
s}=0.944$], whereas the gray solid lines have been obtained by
fixing $n_{\rm s}=0.968$ -- see Eq.~(\ref{nswmap}). We remark that
increasing $|\ks|$'s the required $\kss$'s drop. We observe that
the optimistic constraint of Eq.~(\ref{con}) can be met in a very
limited slice of the allowed area. In this region also $\sgm_*$
turns out to be rather large $(\sim10^{17}~\GeV)$ and therefore we
observe a mild dependence of our results on $c_{6K}$ (or $\ksss$)
too. Also there is a remarkable augmentation of $\as$ which
saturates the bound of \Eref{aswmap} along the thick black solid
line. Namely, for $n_{\rm s}=0.968$ we find
\beqs\bea \label{rennm1} &&
\{0.05\}~2.1\lesssim{\kp\over10^{-2}}\lesssim2.5,~~\{1.5\}
~4.8\lesssim{-\ks\over10^{-2}}\lesssim6.8 \\
\label{rennm2} &&  7.2~\lesssim
{\kss\over10^{-1}}\lesssim7.55~\{11.2\}~~\mbox{and}~~\{1.5\cdot10^{-3}\}~
0.48\lesssim -{\as\over10^{-2}}\lesssim1,\eea\eeqs
where the limiting values obtained without imposing \Eref{con} are
indicated in curly brackets. In the corresponding region, $\Dex$
ranges between $18$ and $38\%$. As can be deduced from
\sFref{nnmf}{a}, $\Dex$ increases with $|\ks|$ or as $\kss$ drops.
%

\begin{figure}[!t]\vspace*{-.16in}
\hspace*{-.2in}
\begin{minipage}{8in}
\epsfig{file=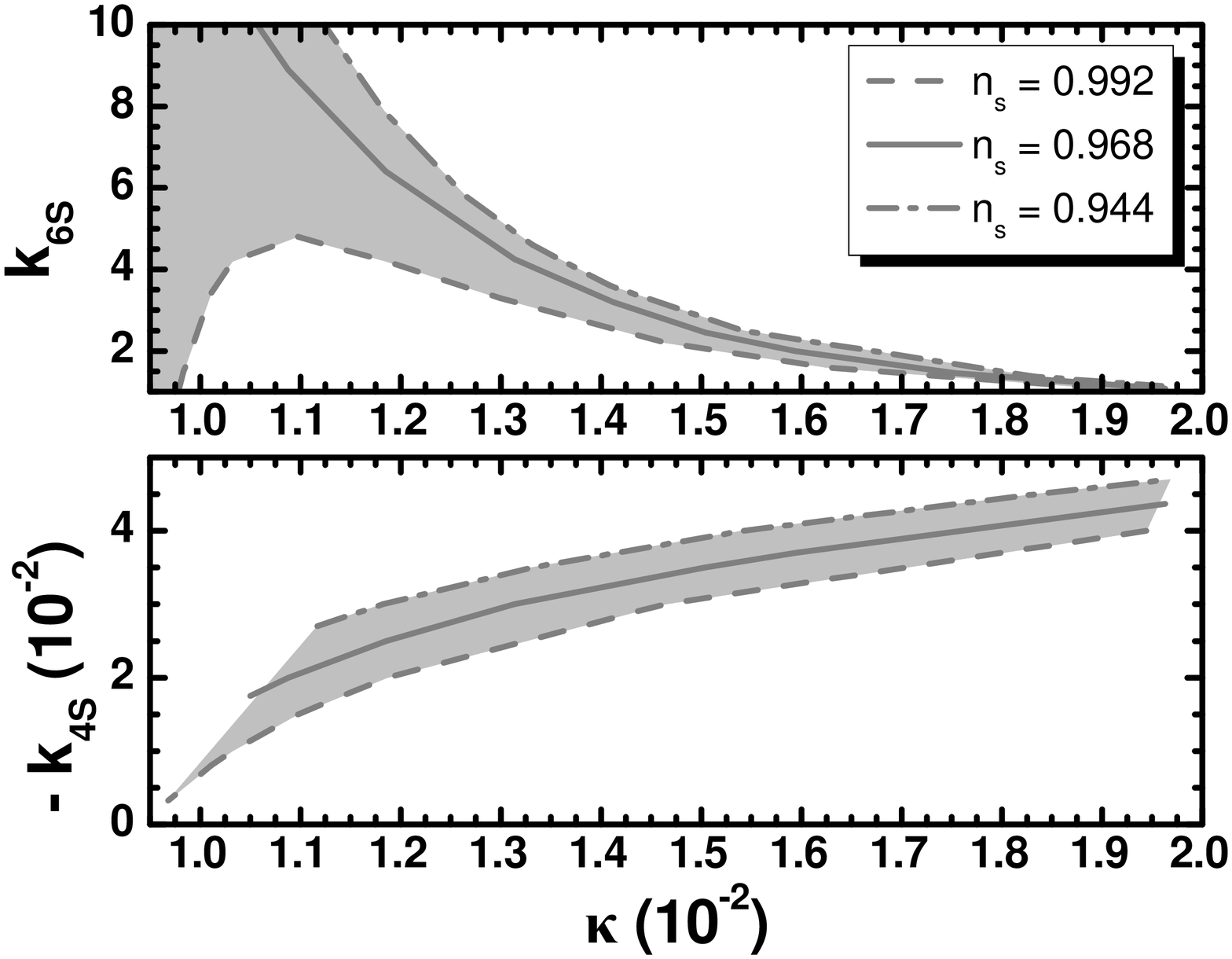,height=3.25in,angle=-90}
\hspace*{-1.25cm}
\epsfig{file=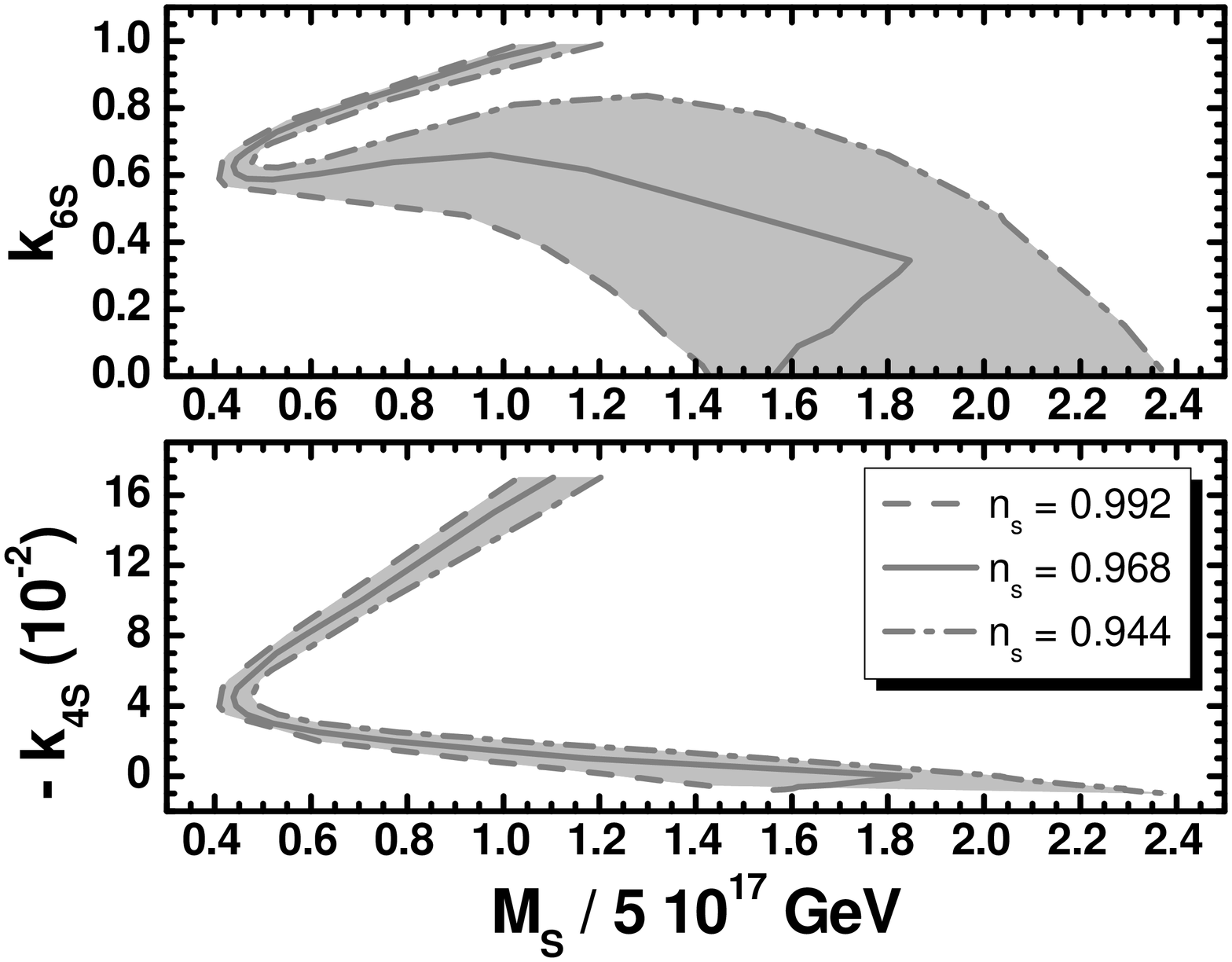,height=3.25in,angle=-90} \hfill
\end{minipage}
\hfill \caption[]{\sl\ftn Allowed (lightly gray shaded) regions,
as determined by Eqs.~(\ref{Nhi})-(\ref{Mgut}), in the
$\kappa-(-\ks)$ and $\kappa-\kss$ planes for shifted FHI (left
graphs) or in the $M_{\rm S}-(-\ks)$ and $M_{\rm S}-\kss$ planes
for smooth FHI (right graphs) in nnmSUGRA. We take
$\ksss=-1.5,\kst=-1$ and $\ksv=0.5$. The conventions adopted for
the various lines are also shown.}\label{nnmfs}
\end{figure}

\subsubsection{\scshape Shifted and Smooth FHI}\label{shnnm}

As in the other scenaria, shifted and smooth FHI can become
consistent with \Eref{Mgut}. Contrary to the the other scenaria,
though, the consideration of two parameters ($\ks$ and $\kss$)
allows here to find wider regions of parameters compatible with
Eqs.~(\ref{Nhi})-(\ref{Mgut}). These are shown in \Fref{nnmfs} for
shifted [smooth] FHI, $\ksss=-1.5,\kst=-1$ and $\ksv=0.5$ (left
[right] panel). The solid, dashed and dot-dashed gray lines
correspond, as usual, to $\ns=0.968,0.992$ and $0.944$
respectively. More specifically,

\begin{itemize}
\item In the case of shifted FHI, we display the allowed regions
in the $\kappa-(-\ks)$ and $\kappa-\kss$ planes. The left and
right boundaries of the allowed regions come from the bounds on
$\xi$ -- see \Sref{Winf}. The results are pretty stable against
variations of $\ksss$ since the used $\kp$'s are rather low
whereas \Eref{con} is violated throughout. For $\ns=0.968$ we
obtain $\Dex\simeq0.29-0.33$ and
$\as\simeq\lf1.2-5.7\rg\cdot10^{-3}$ with
\beqs\beq
1.05\lesssim\kappa/10^{-2}\lesssim1.96,~~1.7\lesssim-\ks/10^{-2}\lesssim4
~~\mbox{and}~~1.1\gtrsim\kss\gtrsim10.\eeq
\item  In the case of smooth FHI, we display the allowed areas in
the $M_{\rm S}-(-\ks)$ and $M_{\rm S}-\kss$ planes. It is worth
mentioning that $\Vhi$ remains monotonic almost in the whole
allowed region depicted in the graphs. Only minor portions close
the dot-dashed line violate \Eref{con}. For $\ns=0.968$ we obtain
$0.6\lesssim|\as|/10^{-3}\lesssim8$ with
\beq0.4\lesssim {M_{\rm
S}/5\cdot10^{17}~\GeV}\lesssim1.82,~~0\lesssim-\ks\lesssim0.17
~~\mbox{and}~~0.3\lesssim\kss \lesssim0.99.\eeq\eeqs

\end{itemize}

In both cases above we observe that $\ks$ is tuned to rather low
values whereas $\kss$ can be adjusted to rather natural values of
order one. This naturalness is removed only in a very minor slice
of the allowed region where positive $\ks$'s can be considered and
the required $\kss$'s are tuned to values of order $0.01$.

\section{\scshape Hilltop FHI in {\bf h}SUGRA}\label{hsugra}

Another, more drastic (and perhaps more radical) way to circumvent
the $n_s$ problem of FHI is the inclusion of extra fields in $K$.
This proposition \cite{mhi} is based on the observation that these
fields provide extra terms in the expressions of $c_{\nu K}$ in
\Eref{Vsugra1}. As a bonus, this construction gives us the
opportunity to elude the notorious $\eta$ problem of FHI. The \Ka
of these extra field is specified in \Eref{hsugra1} and the
structure of the resulting $\Vhi$ is studied in \Sref{hsugra2}.
Our results are presented in \Sref{hsugra3}.

\subsection{\scshape The Relevant Set-up}\label{hsugra1}

As we mention in \Sref{sugra3}, the dependence of $K$ on the
$h_m$'s is encoded in the elements $\K$ and $\Z$. Motivated by
several superstring and D-brane models \cite{Ibanez}, we seek the
following ansatz for them:
\beq \label{K2} \K=\mP^2\sum_{m=1}^{\sf\ftn M}\beta_m
\ln{h_m+h_{m}^{*}\over\mP}\>\>\mbox{and}\>\>\Z=k_Z\prod_{m=1}^{\sf\ftn
M}\left({h_m+h_{m}^{*}\over\mP}\right)^{\alpha_m}\>\>\mbox{with}\>\>\>\bi<0.\eeq
The last restriction in \Eref{K2} is demanded so as to obtain
positivity of the various kinetic energies. We further assume that
$\beta_m$'s have to be integers and $\alpha_m$'s have to be
rational numbers. Also ${\sf\ftn M}$ measures the number of hidden
sector fields. We here restrict ourselves to its lowest possible
value, ${\sf\ftn M}=1$, defining $\alpha:=\al_1$ and
$\beta:=\bt_1$. Note, in passing, that the form used here for $\K$
and $\Z$ has been initially proposed in \cref{nurmi} in order to
justify the saddle point condition needed for the attainment of
$A$-term or MSSM inflation \cite{Aterm}.

In the presence of $\K$ and $\Z$ in \Eref{K2}, the coefficients
$c_{1 K}$ and $c_{2\nu K}$ in the series of \eqs{Vsugra1}{Vol}
receive extra contributions beyond those exposed in
Eqs.~(\ref{c2k})--(\ref{c10k}). The total expressions for $c_{1
K}$ and $c_{2\nu K}$ are found to be
\beqs\bea \label{c1ka} c_{1K}&=&c^{(0)}_{1K}+\al+\bt,\\ \label{c2ka} c_{2K}&=&c^{(0)}_{2K}+{(\al-\bt)^2\over\bt},\\
\label{c4ka} c_{4K}&=&c^{(0)}_{4K}+{(\al-\bt)^3\over\bt^2},\\
\label{c6ka} c_{6K}&=&c^{(0)}_{6K}+ {(\al - \bt)^2 (2 \al^2 - 2
\al \bt + \bt^2)\over2 \bt^3} -\lf \al + {\al^3\over2 \bt^2}
- {5\al^2\over4 \bt} - {\bt\over4}\rg\ks,~~\\
\nonumber c_{8K}&=&c^{(0)}_{8K}+
{(\al - \bt)^2 (6 \al^3 - 6 \al^2 \bt + 3 \al \bt^2 - \bt^3)\over6 \bt^4}\\
&& +\lf{5 \al\over4} - {\al^4\over\bt^3} + {11 \al^3\over4 \bt^2}
- {11 \al^2\over4 \bt} - {\bt\over4}\rg\ks \nonumber\\ &&+\lf{5
\al\over6} +{\al^3\over2 \bt^2} - {7 \al^2\over6 \bt} -
{\bt\over6}\rg\kss, \label{c8ka}\eea \bea \nonumber
c_{10K}&=&c^{(0)}_{10K}+ {(\al - \bt)^2 (24 \al^4 - 24 \al^3 \bt +
12 \al^2 \bt^2 - 4 \al \bt^3 + \bt^4)\over24 \bt^5}\\&& -\lf {3
\al\over4} + {3 \al^5\over2 \bt^4} - {17 \al^4\over4\bt^3} + {9
\al^3\over2 \bt^2} - {19 \al^2\over8 \bt} - {\bt\over8}\rg\ks
\nonumber\\ && -\lf{3 \al\over16} - {\al^4\over4 \bt^3} + {5
\al^3\over8
\bt^2} - {17 \al^2\over32 \bt} - {\bt\over32}\rg\ks^2 \nonumber\\
&& -\lf \al + {\al^4\over\bt^3} - {8 \al^3\over3 \bt^2} + {5
\al^2\over2 \bt} + {\bt\over6}\rg\kss\nonumber\\&& -\lf{3
\al\over4} + {\al^3\over2 \bt^2} - {9 \al^2\over8 \bt} -
{\bt\over8}\rg\ksss. \label{c10ka}\eea\eeqs

\begin{table}[!t]\begin{tabular}[!t]{cc}
\begin{minipage}[t]{6cm}\vspace{-.53in}{\bec
\begin{tabular}{|lll|l||l|}\hline
$-\alpha$&$-\beta$&$\kss$&$\ks$&$-\ckk$\\\hline\hline
$~~4$&$~~3$&$1/6$&$1/3$&$1/3$\\
$9/2$&$~~3$&$1/4$&$3/4$&$1$\\\hline%
$~~4$&$~~2$&$1$&$2$&$2.5$\\
$6/5$ &$~~2$&$1/4$&$8/25$&$0.205$\\ \hline
\end{tabular}\\ \vspace*{0.3cm}
{\sf\small (a)}\eec} \hfill
\end{minipage}
&\begin{minipage}[t]{8cm}
{\bec\begin{tabular}{|lll|l||l|}\hline
$-\alpha$&$-\beta$&$-\kss$&$\ks$&$\ckk$\\\hline\hline
$3$&$~~2$&$~~1/4$&$1/2$&$0$\\
$~~6$&$~~4$&$~~1/2$&$1$&$0$\\\hline%
$~~3$&$~~2$&$~~1/3$&$1/2$&$1/8$\\
$1/2$ &$~~1$&$~~1/4$&$9/4$&$5$\\ \hline
\end{tabular}\\\vspace*{0.3cm}  {\sf\small (b)}\eec}\end{minipage}
\end{tabular}
\hfill \vspace*{-.3in}\caption[]{\sl\ftn  Solutions to \eq{ab} for
$\kss<0$ and $\ck\geq0$ (a) or $\kss>0$ and $\ck<0$
(b).}\label{tabab}
\end{table}

%
From \eqs{Vsugra1}{slow} we infer that a resolution to the $\eta$
problem of FHI requires $\Vhi''=0$ -- needless to say that there
is no contribution to $\eta$ from the term including the $c_{1K}$
coefficient in \eq{Vsugra1}. Consequently, the $\eta$ problem of
FHI can be alleviated, if we demand:
\beq \label{ab} \ck=0\eeq
Moreover, the favored by the data on $\ns$ hilltop FHI can be
attained for $\ckk<0$. Solutions of Eq.~(\ref{ab}) satisfying the
latter restriction are listed in \sTref{tabab}{a}. We observe that
$\kss>0$ is beneficial for the latter result, since it decreases
$\ckk$, without disturbing the fulfilment of \eq{ab}. Another set
of solutions can be taken for $\al=0$. In this case -- which
resembles the cases studied in \cref{sugraP} --, we get
\beq\label{egc4k}
\ckk=3/4,~0,~-3,~-6,-9~~\mbox{for}~~\ks=-\bt=1~~\mbox{and}~~\kss=0,1/2,5/2,9/2,13/2.\eeq
On the other hand, $\ckk\geq0$ is still marginally allowed.
Solutions to \eq{ab} with the latter resulting $\ckk$'s are
arranged in \sTref{tabab}{b}. In both cases, the extracted $\ks$
are confined in the range $0.1-10$, which we consider as being
natural. Note that the realization of FHI in nmSUGRA -- see
\Sref{nmsugra3} -- or nnmSUGRA -- see \Sref{nnmsugra3} -- requires
a significantly lower $|\ks|$, i.e.,
$10^{-3}\lesssim|\ks|\lesssim0.1$.

More generically, taking $\al$, $\bt$ and $\kss$  as input
parameters we can assure the fulfillment of \eq{ab} constraining
$\ks$ via \eq{c2ka} and find $c_{4K}$ through \eq{c4ka}. Working
this way, we plot in \Fref{figab} (left [right] graph) $-\ckk$
[$\ks$] versus $-\alpha$ for $\bt=-2$ and various $\kss$'s (black
points) or $\kss=0.5$ and various $\bt$'s (gray points). The
adopted values for all the free parameters employed are also
shown. From the right plot of \Fref{figab} we remark that the
derived $\ks$'s can be characterized as natural since the majority
of them are of order $1$. These are independent from the used
$\kss$'s since \Eref{c2k} does not contain $\kss$. From the left
plot of \Fref{figab}, we notice that a wide range of negative
$\ckk$'s can be produced which, however, can be bounded from above
by the result $|\ckk|\leq16$. As we show in \Sref{hsugra3}, these
$\ckk$'s can assist us to achieve hilltop FHI consistently with
\Eref{nswmap} for a broad range of $\kappa$'s or $M_{\rm S}$'s.

\begin{figure}[!t]\vspace*{-.16in}
\hspace*{-.2in}
\begin{minipage}{8in}
\epsfig{file=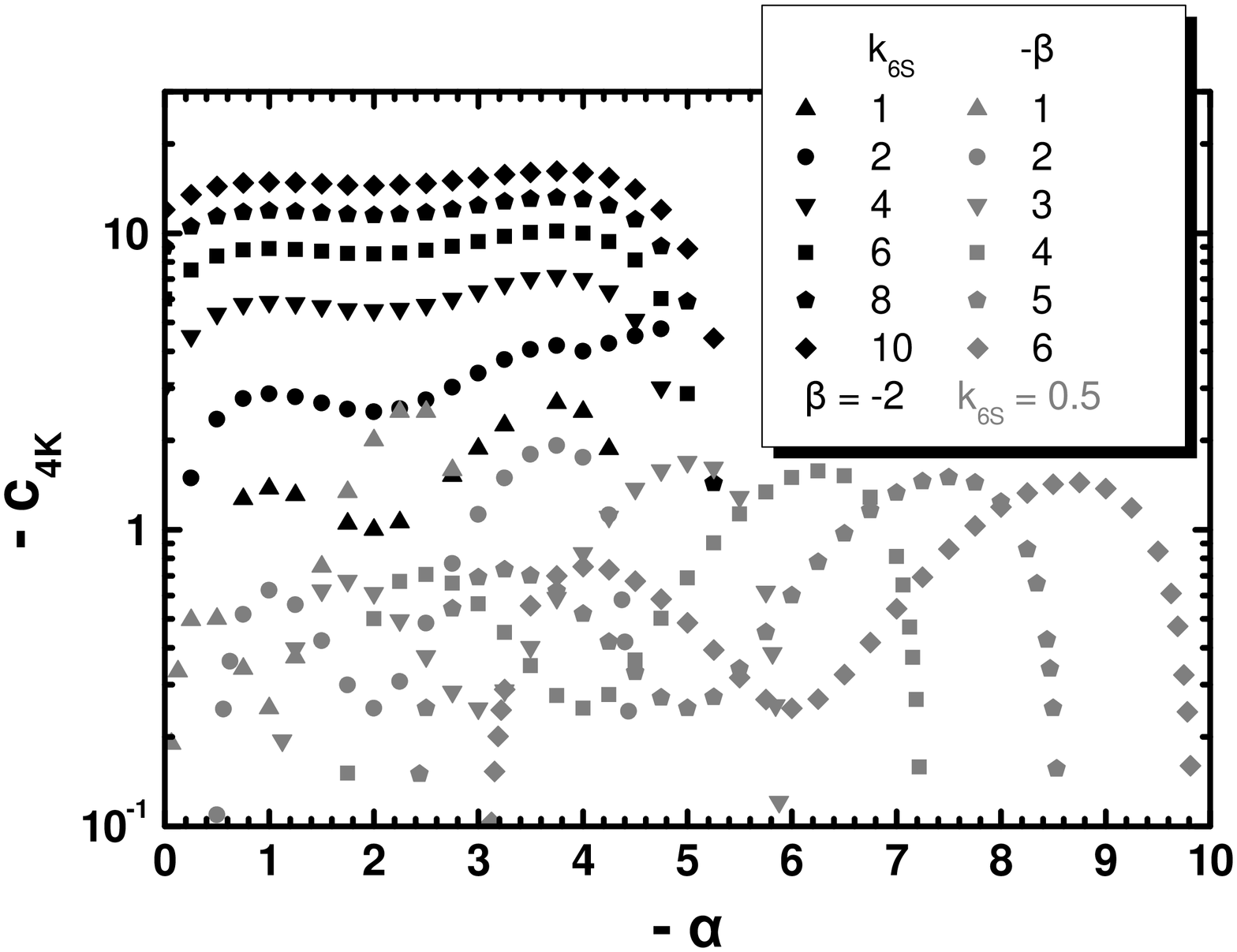,height=3.25in,angle=-90}
\hspace*{-1.25cm}
\epsfig{file=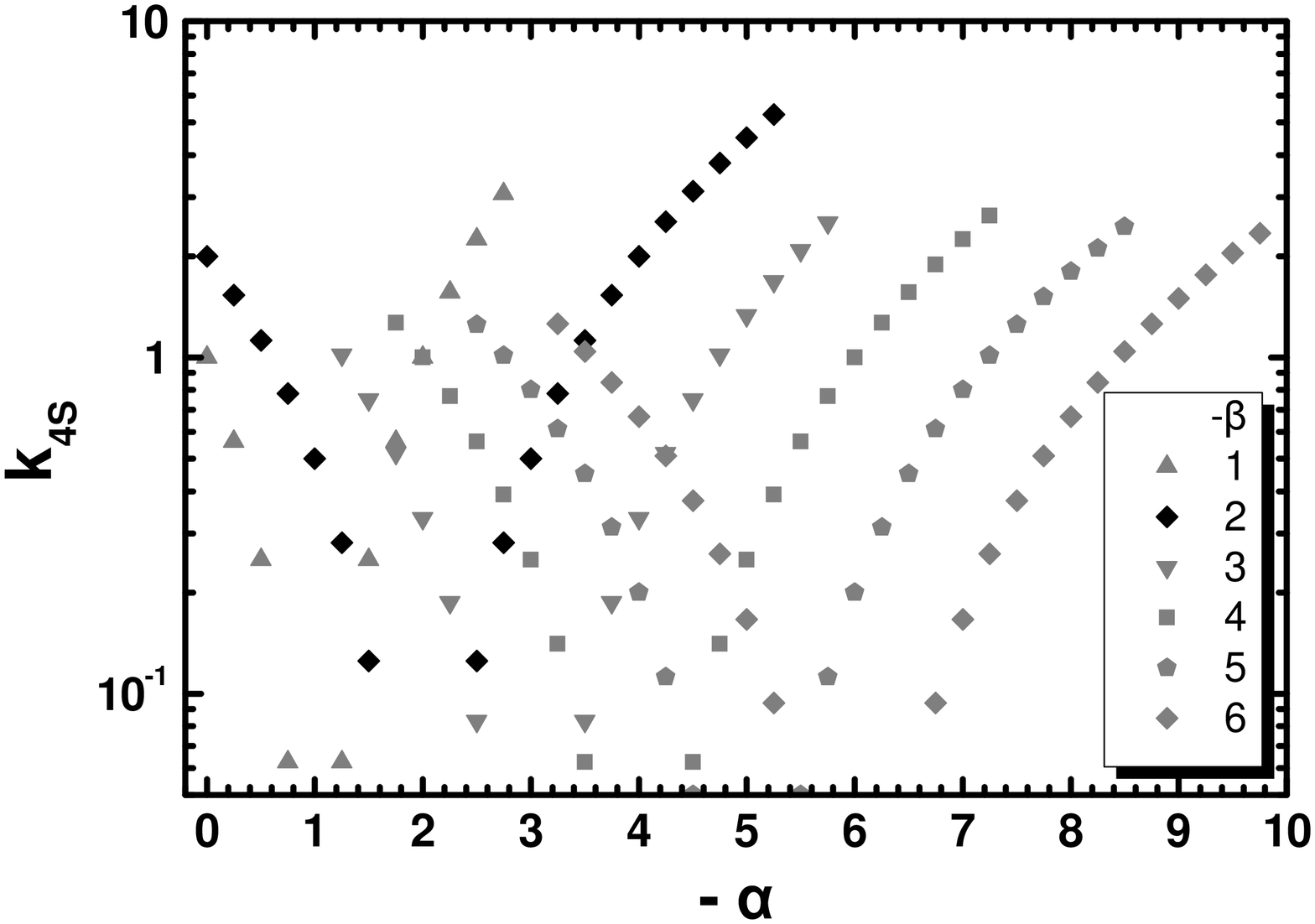,height=3.25in,angle=-90} \hfill
\end{minipage}
\hfill \caption[]{\sl\ftn Values of $-c_{4K}$ obtained from
Eqs.~(53{\sf\ssz c}) and (\ref{ab}) (left plot) and the resulting
$\ks$'s from Eqs.~(53{\sf\ssz b}) and (\ref{ab}) (right plot)
versus $-\alpha$ for $\bt=-2$ and various $\kss$'s (black points)
or $\kss=0.5$ and various $\bt$'s (gray points). The adopted
values for the remaining parameters -- $\kss$ and $\bt$ -- are
also shown.}\label{figab}
\end{figure}

\subsection{\scshape Structure of the Inflationary Potential}
\label{hsugra2}

For $\sgm$ close to $\sgm_*$, $\Vhi$ given in \Eref{Vol} can be
approximated as
\beq\label{Vhsugra} V_{\rm HI}\simeq\Vhio\,\left(1+\ c_{\rm
HI}+\,\ckk{\sigma^4\over4\mP^4}-\ckx{\sigma^6\over8\mP^6}\right),\eeq
where \Eref{ab} is taken into account. As in the case of nnmSUGRA,
a possible ugly runaway behavior of the resulting $\Vhi$ can be
evaded by the inclusion of higher order terms in the expansion of
\Eref{Vsugra1} -- see \Eref{Vol}.

For $\ckk<0$, $V_{\rm HI}$ reaches a maximum at
$\sigma=\sigma_{\rm max}$ which can be estimated as follows:
\begin{equation}
\label{sigmamax3} V'_{\rm HI}(\sigma_{\rm max})=0~~\Rightarrow~~
\sigma_{\rm max}\simeq\left\{\bem
\left(\kappa^2{\sf\ftn
N}/8\pi^2\left|\ckk\right|\right)^{1/4}\hfill & \mbox{for standard
FHI}, \hfill \cr
\left(\kappa^2/4\pi^2\left|\ckk\right|\right)^{1/4}\hfill
&\mbox{for shifted FHI}, \hfill \cr
\left(8\mu_{\rm S}^2M_{\rm
S}^2/27\left|\ckk\right|\right)^{1/8}\hfill &\mbox{for smooth
FHI},
 \hfill \cr\eem
\right.\end{equation}
with $V''_{\rm HI}(\sigma_{\rm max})<0$. Since the behavior of
$\Vhi$ at large $\sigma$'s is dominated by the higher powers of
$\sgm$ in \Eref{Vol}, as in the case of \Sref{nnmsugra}, $\Vhi$
can develop a minimum which is located at $\sgm=\sgm_{\rm min}$
given by \Eref{sigmamin2}.

\begin{figure}[!t]\vspace*{-.16in} \hspace*{-.2in}
\begin{minipage}{8in}
\epsfig{file=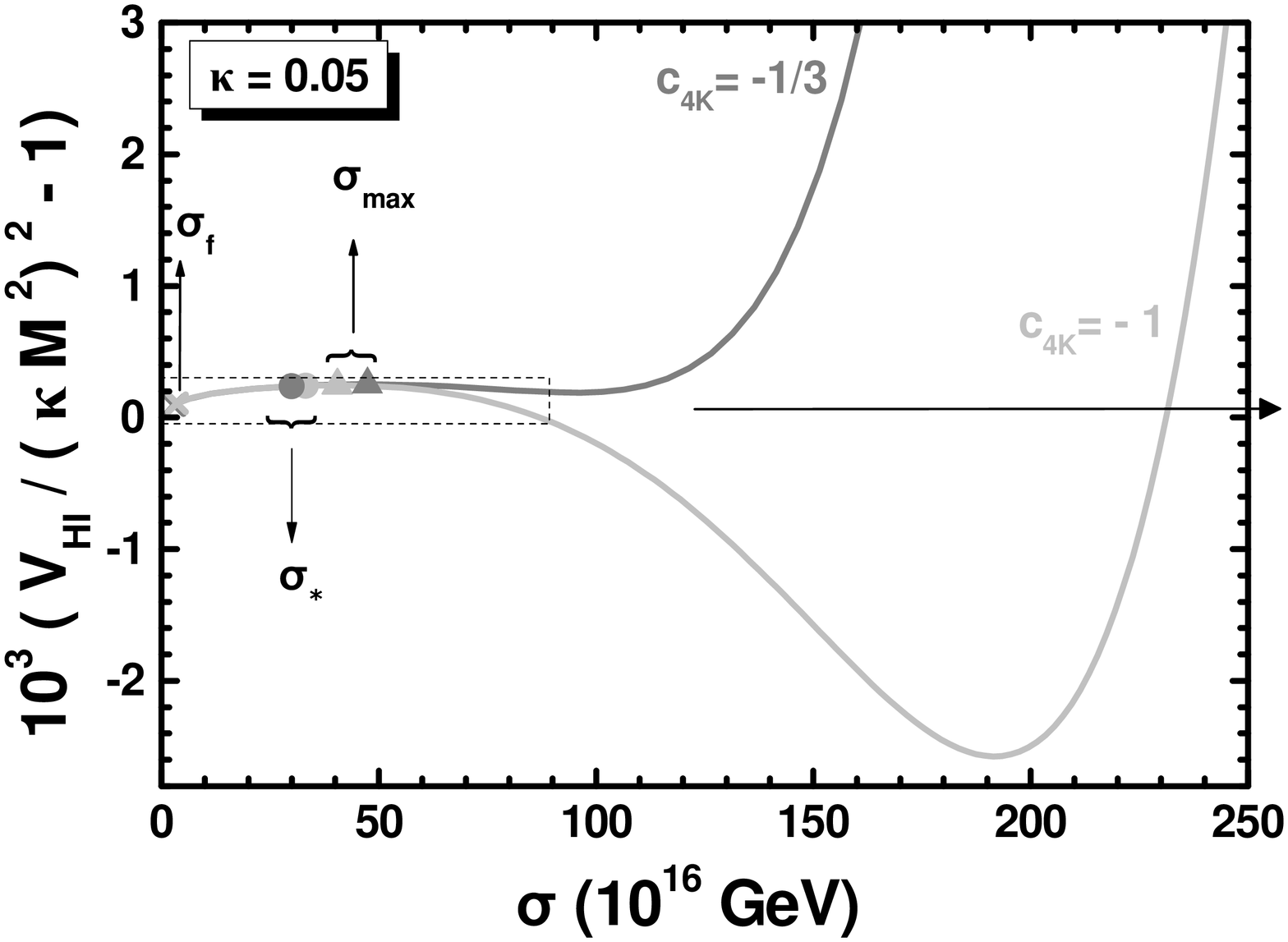,height=3.25in,angle=-90}
\hspace*{-1.25cm}
\epsfig{file=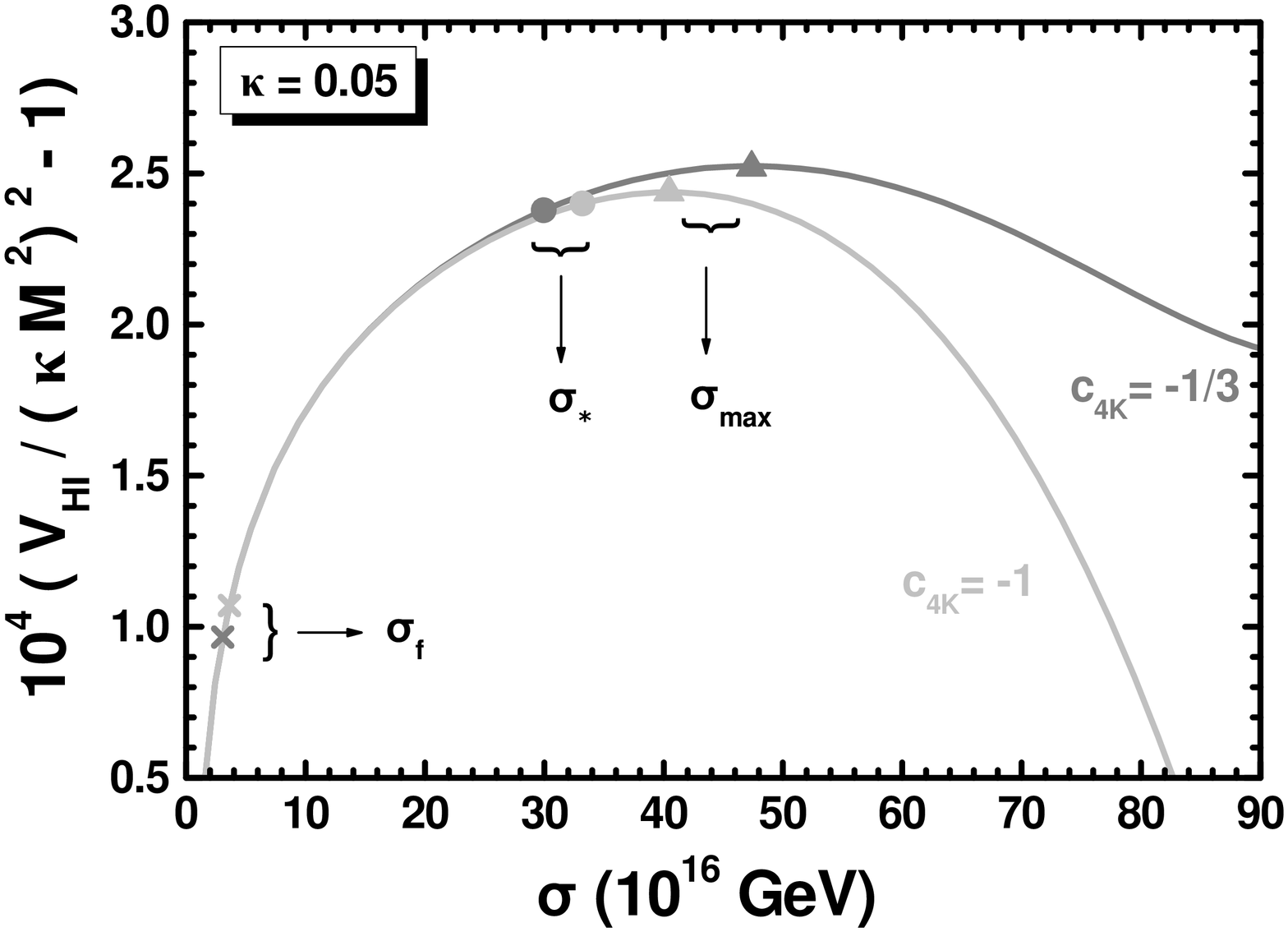,height=3.25in,angle=-90} \hfill
\end{minipage}
\hfill\caption[]{\sl\ftn The variation of $\Vhi$ in \Eref{Vhsugra}
as a function of $\sgm$ for standard FHI in hSUGRA, $\kappa=0.05,
\ksss=-3,\kst=-\ksv=-0.5$ and $\ckk=-1/3$ ($\ns=0.969$) [$\ckk=-1$
($\ns=0.945$)](gray [light gray] line). The values of $\sigma_{\rm
max}$, $\sigma_*$ and $\sigma_{\rm f}$ are also depicted.}
\label{Vhia}
\end{figure}

This structure of $\Vhi$ is visualized in \Fref{Vhia} where we
display its variation as a function of $\sgm$ for $\sgm_{\rm
c}\leq\sgm\leq\mP$ [$\sgm_{\rm c}\leq\sgm\leq2.5\sgm_*$] (left
[right] plot) for standard FHI in hSUGRA,
$\kappa=0.05,~\ksss=-3,~\kst=-\ksv=-0.5$ and $\ckk=-1/3$ (gray
line) or $\ckk=-1$ (light gray line). The parameters $\al,\bt,\ks$
and $\kss$ which lead to the selected $\ckk$'s are shown in
\sTref{tabab}{a}. In the first case (gray line) we obtain
$\ns=0.969$ with $\Dex=0.32$ whereas in the second one, we get
$\ns=0.945$ with $\Dex=0.17$. This result signalizes the presence
of a rather severe tuning needed in order to implement hilltop FHI
as anticipated in \Sref{obs3}. It is also clear that the
minimum-maximum structure of $\Vhi$ remains in both cases with the
second case being much more evident. The values of $\sigma_*$ and
$\sigma_{\rm f}$ are also depicted.

\subsection{\scshape Results} \label{hsugra3}

Our strategy in the numerical investigation of the hSUGRA scenario
is the one described in Sec.~\ref{msugra2}. In addition to the
parameters manipulated there, we have here the parameter $\ckk$
which can be adjusted in order to fulfill Eq.~(\ref{nswmap})
whereas the boundedness of $\Vhi$ is controlled by the $k_{2\nu
S}$'s with $3\leq\nu\leq6$. We finally check if the required
$\ckk$'s can be derived from \eqss{c2ka}{c4ka}{ab} and the
validity of Eq.~(\ref{con}). Our numerical results are presented
in \Sref{sth} for standard FHI and in \Sref{shh} for shifted and
smooth FHI.

We can, however, do some preliminary estimations for the expected
$\ns$'s, following the steps described above \Eref{nssugra}. In
particular we find:
\begin{equation} \label{nssugra3} n_{\rm s}\simeq\left\{\bem
1-{1/N_{\rm HI*}}+{3\kappa^2{\sf\ftn N}N_{\rm HI*}\ckk/4\pi^2}
\hfill & \mbox{for standard FHI}, \hfill \cr
1-{1/N_{\rm HI*}}+{3\kappa^2N_{\rm HI*}\ckk/2\pi^2} \hfill
&\mbox{for shifted FHI}, \hfill \cr
1-{5/3N_{\rm HI*}}+4\ckk\left(6\mu^2_{\rm S}M^2_{\rm S}N_{\rm
HI*}\right)^{1/3}\hfill &\mbox{for smooth FHI}. \hfill \cr\eem
\right.\end{equation}
From the expressions above, we can easily infer that $\ckk<0$ can
diminish significantly $n_{\rm s}$. To this end, in the cases of
standard and shifted FHI, $|\ckk|$ has to be of order unity for
relatively large $\kappa$'s and much larger for lower $\kappa$'s
whereas, for smooth FHI, a rather low $|\ckk|$ is enough.

\begin{figure}[!t]\vspace*{-.16in} \hspace*{-.2in}
\begin{minipage}{8in}
\epsfig{file=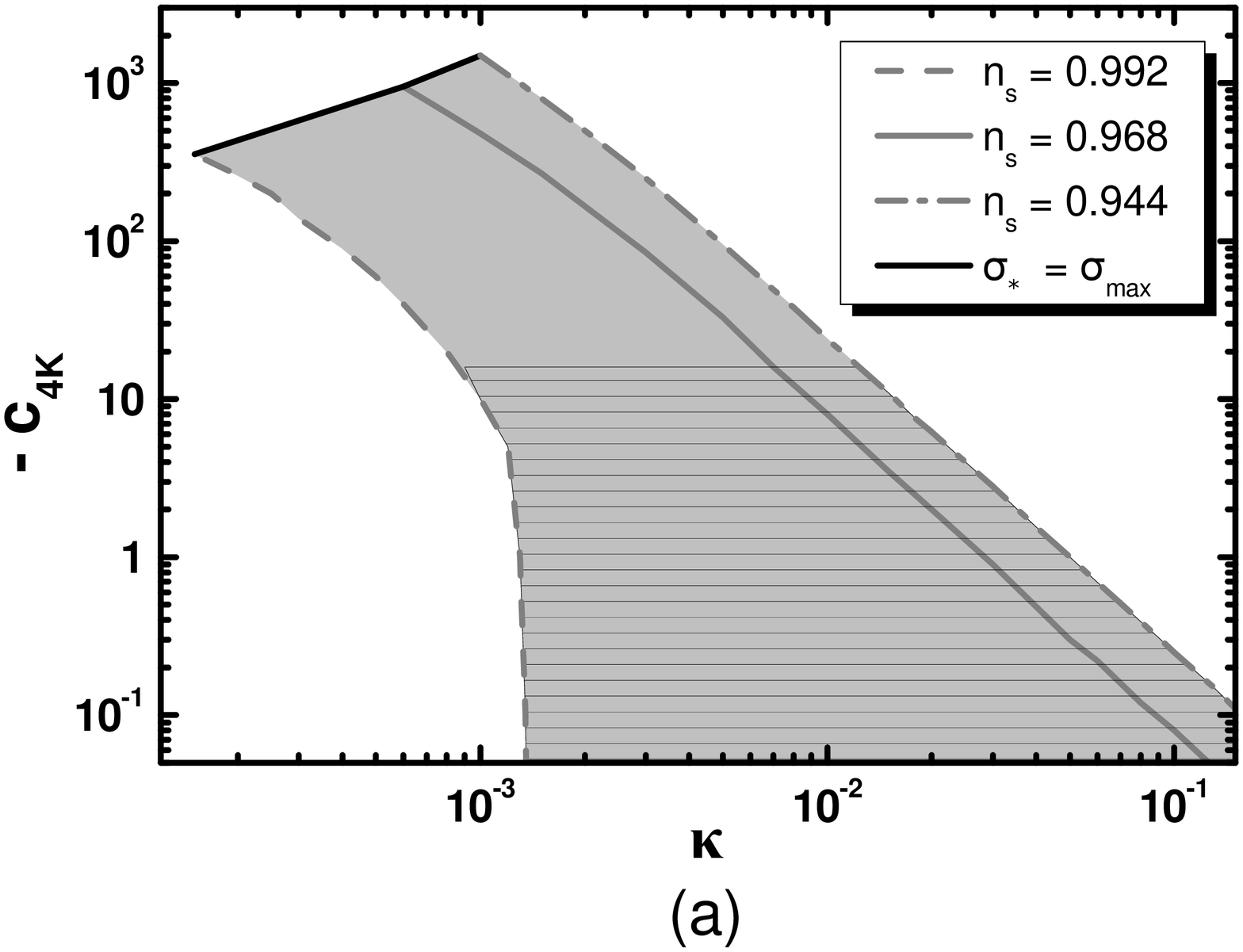,height=3.25in,angle=-90}
\hspace*{-1.25cm}
\epsfig{file=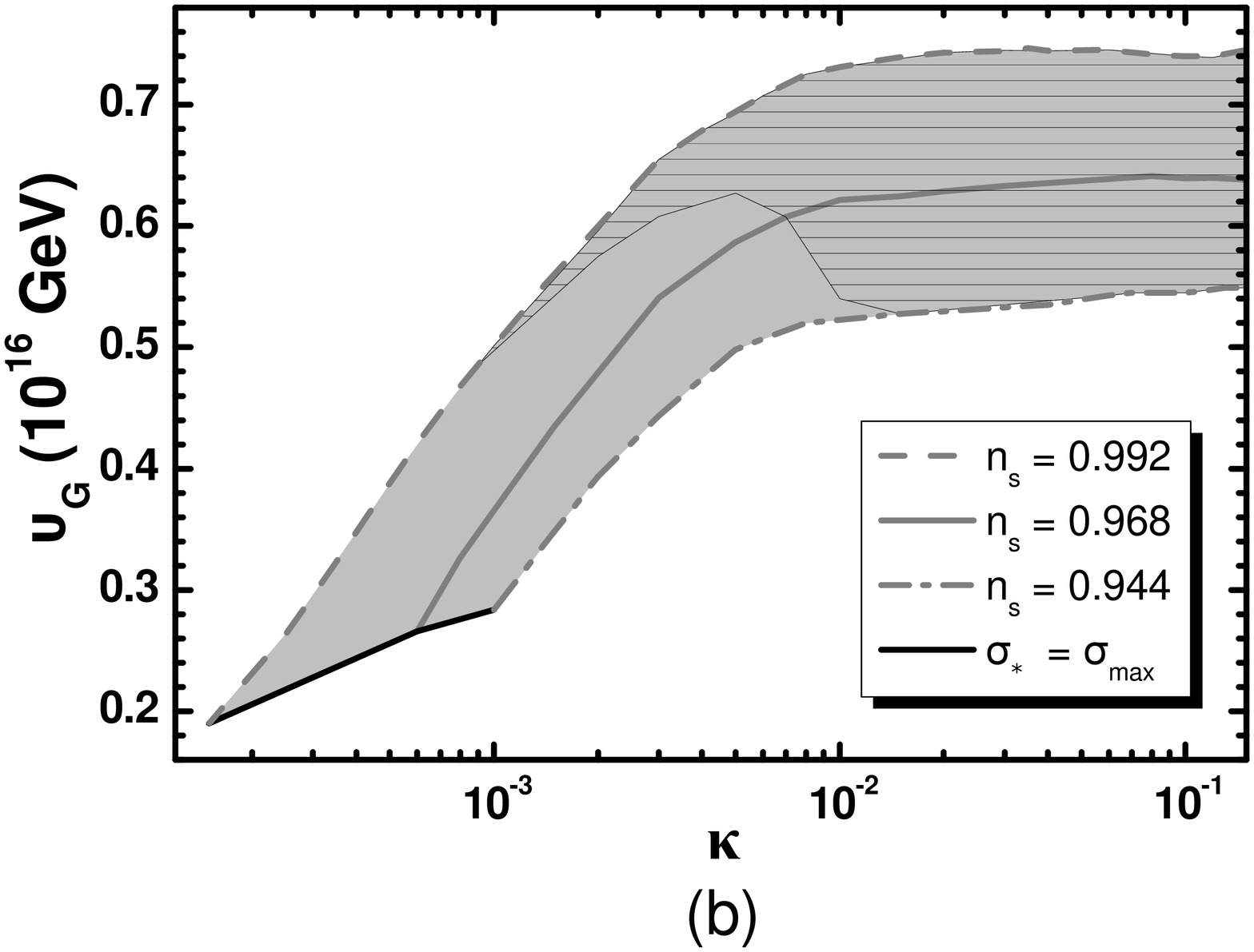,height=3.25in,angle=-90} \hfill
\end{minipage}
\hfill \caption[]{\sl \ftn Allowed (lightly gray shaded) regions,
as determined by Eqs.~(\ref{Nhi})-(\ref{aswmap}), in the
$\kappa-c_{4K}$ [$\kappa-v_{_G}$] plane (a) [(b)] for standard FHI
in hSUGRA with $c_{2K}=0$. Ruled is the region which can be
covered by the values of $\al$, $\bt$ and $\ks$ depicted in
\Fref{figab}. The conventions adopted for the various lines are
also shown. }\label{fig2}
\end{figure}


\subsubsection{\scshape Standard FHI}\label{sth}

In \sFig{fig2}{a} [\sFig{fig2}{b}] we delineate the (lightly gray
shaded) regions allowed by Eqs.~(\ref{Nhi}) -- (\ref{aswmap}), in
the $\kappa-\ckk$ [$\kappa-v_{_G}$] plane for standard FHI. The
conventions adopted for the various lines are also shown in the
r.h.s of each graph. In particular, the black solid [dashed] lines
correspond to $n_{\rm s}=0.992$ [$n_{\rm s}=0.944$], whereas the
gray solid lines have been obtained by fixing $n_{\rm s}=0.968$ --
see Eq.~(\ref{nswmap}). Below the black solid line, our initial
assumption $\sigma_*<\sigma_{\rm max}$ is violated. The various
lines terminate at $\kappa=0.15$, since for larger $\kappa$'s the
two restrictions in \eqs{Nhi}{Prob} cannot be simultaneously met.
Note that for $n_{\rm s}=0.992$ and
$1.3\cdot10^{-3}\lesssim\kappa\lesssim0.15$ the curve is obtained
for positive $0\lesssim\ckk\lesssim0.025$, not displayed in
\sFig{fig2}{a}.

From our data, we can deduce that {\sf\ftn (i)} $v_{_G}$, $\ckk$
and $\Dex$ increase with $\ns$, for fixed $\kappa$ and {\sf\ftn
(ii)} $\ckk$ and $\Dex$ increase with $\kappa$, for fixed $\ns$.
Comparing \sFig{fig2}{a} and \sFref{figab}{a}, we observe that the
required $\ckk$'s, in order to achieve $\ns$'s within the range of
\eq{nswmap}, can be derived from the fundamental parameters of the
proposed K\"ahler potentials -- see \eqs{K}{K2} -- in a wide range
of parameters which is depicted as hatched portions of the light
gray areas in \Fref{fig2}. In particular, for $n_{\rm s}=0.968$ we
obtain
\beqs\bea &&
0.7\lesssim{\kappa\over10^{-2}}\lesssim15,\>\>6.1\lesssim
{\vg\over10^{15}~\GeV}\lesssim6.4,\\&& 16\gtrsim
-\ckk\gtrsim0.035\>\>\mbox{and}\>\>0.30\lesssim\Dex\lesssim0.33.\eea\eeqs
Note that the $\vg$'s encountered here are lower than those
required by \Eref{Mgut}. In this case also, as for nmSUGRA and
nnmSUGRA -- see Secs~\ref{stnm} and \ref{stnnm} --, a certain
degree of tuning is required as can been seen from the values of
$\Dex$ above.

\renewcommand{\arraystretch}{1.2}
\begin{table}[!t]
\begin{center}
\begin{tabular}{|l|lll||l|lll|}
\hline
\multicolumn{4}{|c||}{\sc Shifted FHI}&\multicolumn{4}{|c|}{\sc
Smooth FHI}\\ \hline\hline
$\ckk$ &${-6}$&$-2.5$&${3}$&$\ckk$&${-0.205}$ & ${0}$&${1/8}$\\
$\Dex/10^{-1}$ &  $3.9$&$5$&$-$&$\Dex/10^{-2}$ & $8.3$&$-$&$-$
\\ \hline\hline
$\sigma_*/10^{16}~\GeV$ &$3.8$ &$3.7$&$3.8$&$\sigma_*/10^{16}~\GeV$ & $25.6$&$27$&$28.1$\\
$\kappa/10^{-3}$ & $9$&$9.1$&$9.3$&$M_{\rm S}/5\cdot10^{17}~\GeV$
& $1.97$&$1.5$&$1.3$\\ \hline
$M/10^{16}~\GeV$&$2.28$& $2.29$&$2.3$&$\mu_{\rm S}/10^{16}~\GeV$& $0.083$&$0.11$&$0.13$\\
$1/\xi$ &  $4.32$&$4.35$&$4.4$&$\sigma_{\rm
f}/10^{16}~\GeV$&$13.4$&$13.4$&$13.4$\\
$N_{\rm HI*}$ &  $52.2$&$52.2$&$52.3$&$N_{\rm HI*}$ &
$52.4$&$52.6$&$52.7$\\\hline
$n_{\rm s}$ &  $0.973$&$0.978$&$0.986$&$n_{\rm s}$ & $0.946$&$0.974$&$0.991$\\
$-\alpha_{\rm s}/10^{-4}$ &  $1.9$&$2.6$&$4.3$ &$-\alpha_{\rm
s}/10^{-4}$ & $4.7$&$6.5$&$8.1$\\ \hline
\end{tabular}
\end{center}
\caption {\sl\ftn Input and output parameters consistent with
Eqs.~(\ref{Nhi})-(\ref{Mgut}) for shifted (with $M_{\rm
S}=5\cdot10^{17}~\GeV$) or smooth FHI in hSUGRA and selected
$\ckk$'s indicated in \Tref{tabab}. To ensure the boundedness of
$\Vhi$ in the case of shifted [smooth] FHI we take
$\ksh=-15,\kst=-8$ and $\ksv=-0.5$ [$\ksh=-1$ and
$\kst=-\ksv=-0.5$].}\label{table3}
\end{table}

\subsubsection{\scshape Shifted and Smooth FHI} \label{shh}

In the cases of shifted and smooth FHI, the achievement of
\Eref{Mgut} is possible and so, we can confine ourselves to
solutions consistent with Eqs.~(\ref{Nhi}) -- (\ref{Mgut}) in
Table~\ref{table3}. The selected $\ckk$'s here can be generated by
the initial parameters ($\al,\bt,\ks$ and $\kss$) of our model as
shown in \sTref{tabab}{a} and \Eref{egc4k} for $\ckk<0$ and
\sTref{tabab}{b} for $\ckk\geq0$. The entries without a value
assigned for $\Dex$ refer to cases in which $V_{\rm HI}$ has no
distinguishable maximum. From the data collected in \Tref{table3}
we observe the following:

\begin{itemize}

\item In the case of shifted FHI, the required $\kappa$'s for
fulfilling \Eref{Mgut} come out to be rather low and so, the
reduction of $n_{\rm s}$ to the level dictated by
Eq.~(\ref{nswmap}) requires rather high $\ckk$'s which in turn ask
for large $\kss$'s too. As a consequence, the boundedness of
$\Vhi$ is affected since $-\ckx$ in \Eref{Vol} becomes negative
and rather large $\ksh$'s, $\kst$'s and $\ksv$'s (we here pose
$\ksh=-15,\kst=-8$ and $\ksv=-0.5$). The lowest possible $\ns$
achieved with bounded $\Vhi$ from below is $0.973$ which lies
within the $68\%$ c.l. observationally allowed margin -- see
\Eref{nswmap}.

\item In the case of smooth FHI, $n_{\rm s}$ turns out to be quite
close to its central value in \eq{nswmap} even with $\ckk=0$.
Therefore, in order to reach the central and the lowest value of
$\ns$ in \eq{nswmap}, one needs rather small $\ckk$'s, which may
be obtained from our initial parameters -- see \sFref{figab}{a}.
However, the resulting $\Dex$'s are lower than those of shifted
FHI. On the other hand, the boundedness of $\Vhi$ is not disturbed
here and can be assured for natural values of $\ksh,\kst$ and
$\ksv$ (we use $\ksh=-1$ and $\kst=-\ksv=-0.5$).

\end{itemize}

\newpage

\section{\scshape Conclusions \label{sec:con}}

We reviewed the basic types (standard, shifted and smooth) of FHI
employing four possible embeddings in SUGRA. Each of these can be
characterized by the adopted \Kap. In our work, we considered a
quite generic \Ka in \Eref{K}, from which the various SUGRA
scenaria can be deduced. In particular, the \Ka of mSUGRA, nmSUGRA
and nnmSUGRA can be determined by \Eref{K} substituting
Eqs.~(\ref{mdef}), (\ref{nmdef}) and (\ref{nnmdef}) respectively,
whereas the one of hSUGRA can be derived by \eqs{K}{K2}. A crucial
difference between hSUGRA and the other scenaria is that in hSUGRA
we have taken into account contributions to inflationary
potential, $\Vhi$ in \Eref{Vol}, originating by extra
(hidden-sector) fields obeying a string-inspired \Kap. The
resulting forms of $\Vhi$ implementing FHI in mSUGRA, nmSUGRA,
nnmSUGRA and hSUGRA are written in Eqs.~(\ref{Vmsugra}),
(\ref{Vnm}), (\ref{Vnnm}) and (\ref{Vhsugra}) respectively.

We confronted the considered models of FHI with a number of
observational data and theoretical requirements which are {\sf\ftn
(i)} the need for a solution to the horizon and flatness problems
of the SSB cosmology -- \Eref{Nhi} {\sf\ftn (ii)} the constraints
on $\Delta_{\cal R}$, $\ns$ and $\as$ as result fitting the WMAP7
data by the $\Lambda$CDM model -- see \eqss{Prob}{nswmap}{aswmap};
{\sf\ftn (iii)} the grand unification of the gauge coupling
constants -- \Eref{Mgut}; {\sf\ftn (iv)} the boundedness, the
convergence and the monotonicity of $\Vhi$ -- see \Eref{con}. Our
findings can be summarized as follows:

\begin{itemize}

\item FHI in mSUGRA: The predicted $n_{\rm s}$ is just marginally
consistent with the observational data, since \Eref{nswmap} is
fulfilled either beyond its $68\%$ c.l. or for rather tuned values
of $\kp$ close to $10^{-5}$ due to the presence of the tadpole
term in $\Vhi$ with $\aS\sim1~\TeV$. \Eref{Mgut} can be met only
for shifted and smooth FHI.

\item FHI in nmSUGRA: Acceptable $n_{\rm s}$'s can be obtained by
restricting the parameter $\ks>0$ involved in $\Vhi$ to rather low
values (of order $10^{-3}$). Enforcing the validity of \Eref{con},
the reduction of $n_{\rm s}$ below around $0.95$ is prevented. The
status of \Eref{Mgut} is as in mSUGRA.

\item FHI in nnmSUGRA: By constraining the parameters $\ks<0$ and
$\kss>0$ of $\Vhi$ to values of order $10^{-2}$ and $10^{-1}$
respectively, we can achieve $n_{\rm s}$'s compatible with data.
\Eref{Mgut} can be attained in all the models of FHI. We remark a
sizable enhancement of $|\as|$ whereas $r$ remains well below its
WMAP7 upper bound in \cite{wmap}.

\item FHI in hSUGRA: The two extra parameters ($\al$ and $\bt$)
contained in the \Ka of the extra fields give us the chance to
eliminate the mass-squared term from $\Vhi$ for natural $\ks$'s
and $\kss$'s and generate suitable $\ckk$'s aiming at reducing
$\ns$ to an acceptable level. \Eref{con} can be satisfied only if
$\ckk\geq0$ which gives observationally less interesting $\ns$'s.
\Eref{Mgut} can be met as in mSUGRA.

\end{itemize}

\renewcommand{\arraystretch}{1.}
\begin{table}[!t]
\begin{center}
\begin{tabular}{|l|c|c|c|}\hline
{\sc  Requirements}&\multicolumn{3}{|c|}{\sc Types of FHI}\\
\cline{2-4}&{\sc Standard}&{\sc Shifted}&{\sc
Smooth}\\\hline\hline
&\multicolumn{3}{|c|}{\sc {\rm m}SUGRA -- {\rm see
\Eref{Vmsugra}}}
\\\hline
\Eref{nswmap} &$\sim$&$\sim$&$\times$\\
\Eref{Mgut} &  $\times$&$\surd$&$\surd$\\
\Eref{con} &$\surd$ &$\surd$&$\surd$\\\hline\hline
&\multicolumn{3}{|c|}{\sc {\rm nm}SUGRA  -- {\rm see \Eref{Vnm}}}\\
\hline
\Eref{nswmap} &$\surd$&$\surd$&$\surd$\\
\Eref{Mgut} &  $\times$&$\surd$&$\surd$\\
\Eref{con} &$\sim$ &$\sim$&$\surd$\\
{$0.1\leq|\ks|\leq10$}&$\times$ & $\times$&$\times$\\\hline\hline
&\multicolumn{3}{|c|}{\sc {\rm nnm}SUGRA   -- {\rm see
\Eref{Vnnm}}}\\ \hline
\Eref{nswmap} &$\surd$&$\surd$&$\surd$\\
\Eref{Mgut} &  $\surd$&$\surd$&$\surd$\\
\Eref{con} &$\sim$ &$\times$&$\surd$\\
{$0.1\leq|\ks|\leq10$}&$\times$ & $\times$&$\sim$\\\hline\hline
&\multicolumn{3}{|c|}{\sc {\rm h}SUGRA -- {\rm see
\Eref{Vhsugra}}}\\ \hline
\Eref{nswmap} &$\surd$&$\sim$&$\surd$\\
\Eref{Mgut} &  $\times$&$\surd$&$\surd$\\
\Eref{con} &$\times$ &$\times$&$\sim$\\
{$0.1\leq|\ks|\leq10$}&$\surd$ & $\surd$&$\surd$\\
\hline
\end{tabular}
\end{center}
\caption {\sl\ftn Test performance of the studied models of FHI in
SUGRA. The symbol $\surd$ [$\times$] denotes that the
corresponding requirement is [is not] satisfied, whereas the
symbol $\sim$ stands for a partial or less natural fulfilment of
the requirement.}\label{table4}
\end{table}

Given that from the imposed requirements \eqss{Nhi}{Prob}{aswmap},
the convergence and the boundedness of $\Vhi$ are fulfilled by all
the considered settings of FHI, we are left with a subset of
requirements -- hierarchically represented by
\eqss{nswmap}{Mgut}{con} -- which, in conjunction with the
naturalness inequality $0.1\leq|\ks|\leq10$, can be employed in
order to rate the analyzed models. Note that the latter criterion
does not apply in mSUGRA where $\ks=0$ by definition. Also, we do
not include the naturalness of $\kss$ in our test, since it is
more or less assured in both relevant models (nnmSUGRA and hSUGRA)
and so, it does not influence decisively our comparisons. Our
four-point test is displayed schematically in \Tref{table4}. We
respectively use the symbol $\surd$, $\times$ or $\sim$ when the
corresponding requirement is satisfied, is not satisfied or is
partially and/or less naturally satisfied. From our final score,
we can infer that no model can be regarded as totally
satisfactory, since at least one shortcoming is encountered in all
cases. However, smooth FHI in nnmSUGRA or hSUGRA can be qualified
as the most promising model -- no $\times$ is signed. On the other
hand, the most compelling implementation of standard [shifted] FHI
is within nnmSUGRA [nmSUGRA] since just one $\times$ is listed. In
our last statement we take into account that the four criteria are
imposed hierarchically -- e.g., the attainment of \Eref{nswmap} is
considered as more important than the achievement of \Eref{con}.

Throughout our investigation we concentrated on the predictions
derived from the realizations of FHI, assuming that we had
suitable initial conditions for FHI to take place -- see e.g.
\cref{clesse}. For this reason, we paid special attention to the
monotonicity of $\Vhi$, which is crucial for a relatively natural
attainment of FHI. In general, it is not clear \cite{gpp, mur} how
the inflaton can reach the maximum of $\Vhi$ in the context of
hilltop inflation. Probably an era of eternal inflation prior to
FHI could be useful \cite{lofti} in order the proper initial
conditions to be set.

Let us finally note that a complete inflationary scenario should
specify the transition to the radiation dominated era and also
explain the origin of the observed baryon asymmetry. For FHI in
mSUGRA or nmSUGRA this has been extensively studied -- see, e.g.,
Ref.~\cite{jean, pana1, lept}. Obviously our models preserve many
of these successful features of this post-inflationary evolution
which may further constrain their parameter space and help us to
distinguish the most compelling version of FHI. Moreover, the
proposed scenaria will be even more challenged by the measurements
of the Planck satellite \cite{planck} which is expected to give
results on $n_{\rm s}$ with an accuracy $\Delta n_{\rm s}\simeq
0.01$ by the next spring.

\subsection*{\scshape Acknowledgments} The work of R.A. was supported
by the Tomalla Foundation. We would like to thank G.~Lazarides for
helpful discussions, S.~Clesse, D.H. Lyth and D.~Nolde for
interesting suggestions.


\def\ijmp#1#2#3{{\sl Int. Jour. Mod. Phys.}
{\bf #1},~#3~(#2)}
\def\plb#1#2#3{{\sl Phys. Lett. B }{\bf #1},~#3~(#2)}
\def\zpc#1#2#3{{Z. Phys. C }{\bf #1},~#3~(#2)}
\def\prl#1#2#3{{\sl Phys. Rev. Lett.}
{\bf #1},~#3~(#2)}
\def\rmp#1#2#3{{Rev. Mod. Phys.}
{\bf #1},~#3~(#2)}
\def\prep#1#2#3{{\sl Phys. Rep. }{\bf #1},~#3~(#2)}
\def\prd#1#2#3{{\sl Phys. Rev. D }{\bf #1},~#3~(#2)}
\def\npb#1#2#3{{\sl Nucl. Phys. }{\bf B#1},~#3~(#2)}
\def\npps#1#2#3{{Nucl. Phys. B (Proc. Sup.)}
{\bf #1},~#3~(#2)}
\def\mpl#1#2#3{{Mod. Phys. Lett.}
{\bf #1},~#3~(#2)}
\def\arnps#1#2#3{{Annu. Rev. Nucl. Part. Sci.}
{\bf #1},~#3~(#2)}
\def\sjnp#1#2#3{{Sov. J. Nucl. Phys.}
{\bf #1},~#3~(#2)}
\def\jetp#1#2#3{{JETP Lett. }{\bf #1},~#3~(#2)}
\def\app#1#2#3{{Acta Phys. Polon.}
{\bf #1},~#3~(#2)}
\def\rnc#1#2#3{{Riv. Nuovo Cim.}
{\bf #1},~#3~(#2)}
\def\ap#1#2#3{{Ann. Phys. }{\bf #1},~#3~(#2)}
\def\ptp#1#2#3{{Prog. Theor. Phys.}
{\bf #1},~#3~(#2)}
\def\apjl#1#2#3{{Astrophys. J. Lett.}
{\bf #1},~#3~(#2)}
\def\n#1#2#3{{Nature }{\bf #1},~#3~(#2)}
\def\apj#1#2#3{{Astrophys. J.}
{\bf #1},~#3~(#2)}
\def\anj#1#2#3{{Astron. J. }{\bf #1},~#3~(#2)}
\def\mnras#1#2#3{{MNRAS }{\bf #1},~#3~(#2)}
\def\grg#1#2#3{{Gen. Rel. Grav.}
{\bf #1},~#3~(#2)}
\def\s#1#2#3{{Science }{\bf #1},~#3~(#2)}
\def\baas#1#2#3{{Bull. Am. Astron. Soc.}
{\bf #1},~#3~(#2)}
\def\ibid#1#2#3{{\it ibid. }{\bf #1},~#3~(#2)}
\def\cpc#1#2#3{{Comput. Phys. Commun.}
{\bf #1},~#3~(#2)}
\def\astp#1#2#3{{\sl Astropart. Phys.}
{\bf #1},~#3~(#2)}
\def\epjc#1#2#3{{Eur. Phys. J. C}
{\bf #1},~#3~(#2)}
\def\nima#1#2#3{{Nucl. Instrum. Meth. A}
{\bf #1},~#3~(#2)}
\def\jhep#1#2#3{{\sl J. High Energy Phys.}
{\bf #1},~#3~(#2)}
\def\jcap#1#2#3{{\sl J. Cosmol. Astropart. Phys.}
{\bf #1},~#3~(#2)}

\newcommand{\hepph}[1]{{\tt hep-ph/#1}}
\newcommand{\hepth}[1]{{\tt hep-th/#1}}
\newcommand{\hepex}[1]{{\tt hep-ex/#1}}
\newcommand{\astroph}[1]{{\tt astro-ph/#1}}
\newcommand{\arxiv}[1]{{\tt arXiv:#1}}

\end{document}